\documentclass[10pt,conference]{IEEEtran}

\usepackage{cite}
\usepackage{amsmath,amssymb,amsfonts}
\usepackage{algorithmic}
\usepackage{graphicx}
\usepackage{textcomp}
\usepackage{xcolor}
\usepackage[hyphens]{url}
\usepackage{fancyhdr}
\usepackage{hyperref}

\usepackage{booktabs}
\usepackage{multirow}
\usepackage{multicol}
\usepackage{pifont}
\usepackage{mathptmx}
\usepackage[normalem]{ulem}
\usepackage[final]{microtype}
\usepackage{algorithm}
\usepackage{dashrule}
\usepackage{stfloats}
\usepackage{xtab}
\usepackage{footnote}
\usepackage{longtable}
\usepackage{tablefootnote}
\usepackage[most]{tcolorbox}
\newtcolorbox{insightbox}{
    sharpish corners, 
    boxrule = 0pt,
    toprule = 1.5pt, 
    bottomrule = 1.5pt, 
    enhanced,
    fuzzy shadow = {0pt}{-2pt}{-0.5pt}{0.5pt}{black!35} 
}

\makesavenoteenv{tabular}
\makesavenoteenv{table}
\usepackage{tabularx}
\usepackage{tabu}
\usepackage{array}
\newcolumntype{P}[1]{>{\centering\arraybackslash}p{#1}}

\definecolor{darkgray}{rgb}{0.3, 0.3, 0.3}
\definecolor{darkgreen}{rgb}{0.0, 0.7, 0.0}
\definecolor{darkyellow}{rgb}{0.8, 0.6, 0.0}
\definecolor{darkred}{rgb}{0.7, 0.0, 0.0}

\newcommand{\darkyellowcheck}{{\color{darkyellow}\ding{51}}}
\newcommand{\darkgreencheck}{{\color{darkgreen}\ding{51}}}
\newcommand{\darkredcross}
{{\color{darkred}\ding{55}}}

\newcommand{\OH}[1]{{\color{violet}{[}\textbf{OH: #1}{]}}}

\newcommand{\hpcayear}{}

\newcommand{\hpcasubmissionnumber}{32}
\title{DFModel: Design Space Optimization of Large-Scale Systems Exploiting Dataflow Mappings}

\def\hpcacameraready{} 

\newcommand\hpcaauthors{Sho Ko, Nathan Zhang, Olivia Hsu, Ardavan Pedram, Kunle Olukotun}
\newcommand\hpcaaffiliation{Stanford University}
\newcommand\hpcaemail{\texttt{\{kosho, stanfurd, owhsu, perdavan, kunle\}@stanford.edu}}

\author{
  \ifdefined\hpcacameraready
    \IEEEauthorblockN{\hpcaauthors{}}
      \IEEEauthorblockA{
        \hpcaaffiliation{} \\
        \hpcaemail{}
      }
  \else
    \IEEEauthorblockN{\normalsize{HPCA \hpcayear{} Submission
      \textbf{\#\hpcasubmissionnumber{}}} \\
      \IEEEauthorblockA{
        Confidential Draft \\
        Do NOT Distribute!!
      }
    }
  \fi 
}

\fancypagestyle{camerareadyfirstpage}{%
  \fancyhead{}
  
  \fancyhead[C]{
    \ifdefined\aeopen
    \parbox[][12mm][t]{13.5cm}{\hpcayear{} IEEE International Symposium on High-Performance Computer Architecture (HPCA)}    
    \else
      \ifdefined\aereviewed
      \parbox[][12mm][t]{13.5cm}{\hpcayear{} IEEE International Symposium on High-Performance Computer Architecture (HPCA)}
      \else
      \ifdefined\aereproduced
      \parbox[][12mm][t]{13.5cm}{\hpcayear{} IEEE International Symposium on High-Performance Computer Architecture (HPCA)}
      \else
      \parbox[][0mm][t]{13.5cm}{\hpcayear{}}
    \fi 
    \fi 
    \fi 
    \ifdefined\aeopen 
      \includegraphics[width=12mm,height=12mm]{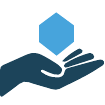}
    \fi 
    \ifdefined\aereviewed
      \includegraphics[width=12mm,height=12mm]{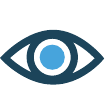}
    \fi 
    \ifdefined\aereproduced
      \includegraphics[width=12mm,height=12mm]{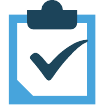}
    \fi
  }
  \fancyfoot[C]{}
}
\begin{document}
\maketitle

\ifdefined\hpcacameraready 
  \thispagestyle{plain}
  \pagestyle{plain}
\else
  \thispagestyle{plain}
  \pagestyle{plain}
\fi

\newcommand{\hpcaheight}{0mm}
\ifdefined\eaopen
\renewcommand{\hpcaheight}{12mm}
\fi


\begin{abstract}
We propose DFModel, a modeling framework for mapping dataflow computation graphs onto large-scale systems.
Mapping a workload to a system requires optimizing dataflow mappings at various levels, including the inter-chip (between chips) level and the intra-chip (within a chip) level.
DFModel is, to the best of our knowledge, the first framework to perform the optimization at multiple levels of the memory hierarchy and the interconnection network hierarchy.
We use DFModel to explore a wide range of workloads on a variety of systems.
Evaluated workloads include two state-of-the-art machine learning applications (Large Language Models and Deep Learning Recommendation Models) and two high-performance computing applications (High Performance LINPACK and Fast Fourier Transform).
System parameters investigated span the combination of dataflow and traditional accelerator architectures, memory technologies (DDR, HBM), interconnect technologies (PCIe, NVLink), and interconnection network topologies (torus, DGX, dragonfly).
For a variety of workloads on a wide range of systems, the DFModel provided a mapping that predicts an average of $1.25\times$ better performance compared to the ones measured on real systems.
DFModel shows that for large language model training, dataflow architectures achieve $1.52\times$ higher performance, $1.59\times$ better cost efficiency, and $1.6\times$ better power efficiency compared to non-dataflow architectures.
On an industrial system with dataflow architectures, the DFModel-optimized dataflow mapping achieves a speedup of $6.13\times$ compared to non-dataflow mappings from previous performance models such as Calculon, and $1.52\times$ compared to a vendor provided dataflow mapping.
The source code is available at Github\footnote{\href{https://github.com/kosho2013/DFModel}{https://github.com/kosho2013/DFModel}}.
\end{abstract}
\section{Introduction}

\newcommand*\colourcheck[1]{%
  \expandafter\newcommand\csname #1check\endcsname{\textcolor{#1}{\ding{52}}}%
}
\newcommand*\colourcross[1]{%
  \expandafter\newcommand\csname #1cross\endcsname{\textcolor{#1}{\ding{56}}}%
}
\colourcheck{green}
\colourcross{red}

\begin{table*}[t!]
\centering

\caption{Comparison between DFModel and previous performance models.\\
*: \darkyellowcheck single kernel, \darkgreencheck multi-kernel fusion.}

\begin{tabular}{lccccccccc}
\toprule
    \multicolumn{1}{c}{\multirow{2}{*}{Models}} & 
    
    \multicolumn{1}{c}{\multirow{2}{*}{\renewcommand{\arraystretch}{1.2} \begin{tabular}{@{}c@{}}General \\ Workloads\end{tabular}}} &

    \multicolumn{1}{c}{\multirow{2}{*}{\renewcommand{\arraystretch}{1.2} \begin{tabular}{@{}c@{}}Compute/Memory/Network \\ Joint Optimization\end{tabular}}} &
    
    \multicolumn{3}{c}{Inter-chip Dataflow} & 
    
    \multicolumn{1}{c}{\multirow{2}{*}{\renewcommand{\arraystretch}{1.2} \begin{tabular}{@{}c@{}}Intra-chip \\ Dataflow*\end{tabular}}} &
    
    \multicolumn{1}{c}{\multirow{2}{*}{\renewcommand{\arraystretch}{1.2} \begin{tabular}{@{}c@{}}On-chip \\ Compute/Memory\end{tabular}}} & 

    \multicolumn{1}{c}{\multirow{2}{*}{\renewcommand{\arraystretch}{1.2} \begin{tabular}{@{}c@{}}Off-chip \\ Memory\end{tabular}}} & 
    
    \multicolumn{1}{c}{\multirow{2}{*}{\renewcommand{\arraystretch}{1.2} \begin{tabular}{@{}c@{}}Off-chip \\
    Network\end{tabular}}}\\
    
    \cmidrule(lr){4-6}
    & & & TP & PP & DP &  &  & \\
    \midrule

    FlexFlow~\cite{jia2019beyond} & \darkredcross & \darkredcross & \darkgreencheck & \darkredcross & \darkgreencheck & \darkredcross & \darkredcross & \darkgreencheck & \darkgreencheck \\
    
    PipeMare~\cite{yang2021pipemare} & \darkredcross & \darkredcross & \darkredcross & \darkgreencheck & \darkredcross & \darkredcross & \darkredcross & \darkgreencheck & \darkredcross \\

    Alpa~\cite{zheng2022alpa} & \darkredcross & \darkredcross & \darkgreencheck & \darkgreencheck & \darkgreencheck & \darkredcross & \darkredcross & \darkgreencheck & \darkgreencheck \\

    Megatron-LM~\cite{narayanan2021efficient} & \darkredcross & \darkredcross & \darkgreencheck & \darkgreencheck & \darkgreencheck & \darkredcross & \darkredcross & \darkgreencheck & \darkredcross \\
    
    Calculon~\cite{isaev2023calculon} & \darkredcross & \darkredcross & \darkgreencheck & \darkgreencheck & \darkgreencheck & \darkredcross & \darkredcross & \darkgreencheck & \darkredcross \\

    LLM-Viewer~\cite{yuan2024llm} & \darkredcross & \darkredcross & \darkredcross & \darkredcross & \darkredcross & \darkgreencheck & \darkredcross & \darkgreencheck & \darkredcross \\

    Rail-Only~\cite{wang2023optimized} & \darkredcross & \darkredcross & \darkgreencheck & \darkgreencheck & \darkgreencheck & \darkredcross & \darkredcross & \darkgreencheck & \darkgreencheck \\

    Timeloop~\cite{parashar2019timeloop} & \darkredcross & \darkredcross & \darkredcross & \darkredcross & \darkredcross & \darkyellowcheck & \darkgreencheck & \darkgreencheck & \darkredcross \\

    FAST~\cite{zhang2022full} & \darkredcross & \darkredcross & \darkredcross & \darkredcross & \darkredcross & \darkgreencheck & \darkgreencheck & \darkgreencheck & \darkredcross \\

    CoSA~\cite{huang2021cosa} & \darkredcross & \darkredcross & \darkredcross & \darkredcross & \darkredcross & \darkyellowcheck & \darkgreencheck & \darkgreencheck & \darkredcross \\
    
    ASTRA-sim~\cite{rashidi2020astra} & \darkredcross & \darkredcross & \darkgreencheck & \darkgreencheck & \darkgreencheck & \darkredcross & \darkredcross & \darkgreencheck & \darkgreencheck \\
    
    Explainable-DSE~\cite{dave2023explainable} & \darkredcross & \darkredcross & \darkredcross & \darkredcross & \darkredcross & \darkyellowcheck & \darkgreencheck & \darkgreencheck & \darkredcross \\

    LLMCompass~\cite{llmcompass} & \darkredcross & \darkredcross & \darkgreencheck & \darkgreencheck & \darkredcross & \darkredcross & \darkgreencheck & \darkgreencheck & \darkredcross \\

    Orojenesis~\cite{mindthegap} & \darkredcross & \darkredcross & \darkredcross & \darkredcross & \darkredcross & \darkgreencheck & \darkgreencheck & \darkgreencheck & \darkredcross \\

    GenZ~\cite{bambhaniya2024demystifying} & \darkredcross & \darkredcross & \darkgreencheck & \darkgreencheck & \darkredcross & \darkredcross & \darkredcross & \darkgreencheck & \darkgreencheck \\

    TACOS~\cite{won2024tacos} & \darkredcross & \darkredcross & \darkgreencheck & \darkgreencheck & \darkgreencheck & \darkredcross & \darkredcross & \darkredcross & \darkgreencheck \\

    LoopTree~\cite{gilbert2024looptree} & \darkredcross & \darkredcross & \darkredcross & \darkredcross & \darkredcross & \darkgreencheck & \darkgreencheck & \darkgreencheck & \darkredcross \\

    vTrain~\cite{bang2024vtrain} & \darkredcross & \darkredcross & \darkgreencheck & \darkgreencheck & \darkgreencheck & \darkredcross & \darkredcross & \darkgreencheck & \darkgreencheck \\

    AMPeD~\cite{moolchandani2023amped} & \darkredcross & \darkredcross & \darkgreencheck & \darkgreencheck & \darkgreencheck & \darkredcross & \darkredcross & \darkgreencheck & \darkgreencheck \\

    Tale of Two Cs~\cite{pati2023tale} & \darkredcross & \darkredcross & \darkgreencheck & \darkgreencheck & \darkgreencheck & \darkredcross & \darkredcross & \darkgreencheck & \darkgreencheck \\

    Habitat~\cite{geoffrey2021habitat} & \darkredcross & \darkredcross & \darkredcross & \darkredcross & \darkredcross & \darkyellowcheck & \darkgreencheck & \darkgreencheck & \darkredcross \\

    \textbf{DFModel} & \darkgreencheck & \darkgreencheck & \darkgreencheck & \darkgreencheck & \darkgreencheck & \darkgreencheck & \darkgreencheck & \darkgreencheck & \darkgreencheck \\
  
    \bottomrule
\end{tabular}
\label{framework}
\vspace{-10pt}
\end{table*}

\begin{figure*}[t!]
  \centering
  \includegraphics[width=\linewidth]{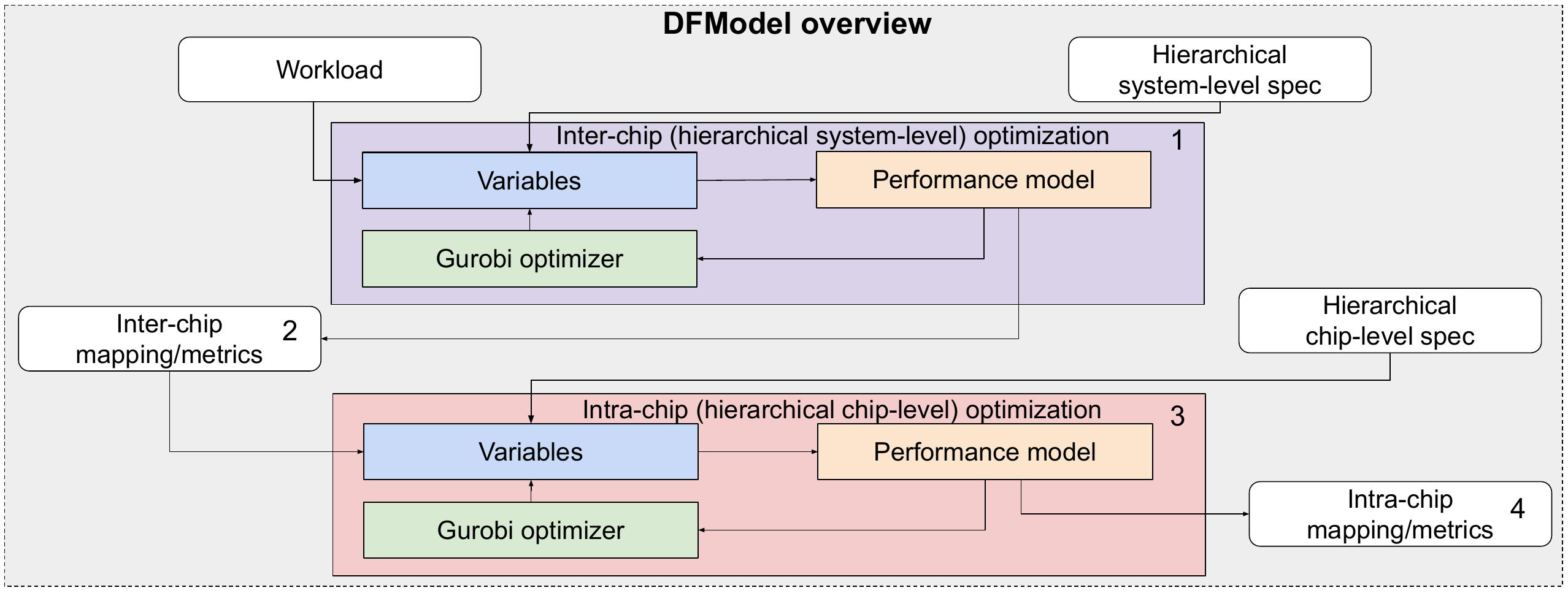}
  \vspace{-20pt}
  \caption{The overview of DFModel.
  DFModel takes in a workload description represented by a dataflow graph and a system specification including a multi-node distributed system and the individual data parallel chip.
  DFModel goes through two layers of optimization: an inter-chip layer (1) for hierarchical system-level optimization and an intra-chip layer (3) for hierarchical chip-level optimization.
  (1) takes in workload and hierarchical system-level specification and produces inter-chip mapping and metrics (2).
  Then (2) and hierarchical chip-level specification are fed into (3) to produce intra-chip mapping and metrics (4).
  We assume a typical chip will be within a region close to pareto-optimal design for the balance between memory and computation similar to existing accelerators, such as GPUs~\cite{choquette2022nvidia}, TPUs~\cite{jouppi2023tpu}, and SambaNova RDUs~\cite{prabhakar2024sambanova}.
  }
  \label{DFModel_figure1}
  \vspace{-20pt}
\end{figure*}

Accurate estimation of the performance of large-scale distributed systems using analytical models continues to be an important and challenging problem.
Over time, distributed system workloads have evolved from traditional high-performance computing (HPC) applications to modern deep learning (DL) applications.
As such, researchers have built analytical models for several of these HPC and DL workloads, such as High Performance LINPACK (HPL)~\cite{kim2022snuhpl}, Fast Fourier Transforms (FFTs)~\cite{ayala2022analysis}, Deep Learning Recommendation Models (DLRMs)~\cite{lin2022building}, and Large Language Models (LLMs)~\cite{isaev2023calculon}.
However, these models are limited to certain classes of applications, and are not general enough for a broad range of applications.
In this paper, we propose a more general modeling framework based on a dataflow representation that can model arbitrary workloads, and can be mapped to arbitrary system specifications with a wide variety of memory and interconnect hierarchies.

\begin{figure}[t!]
  \centering
  \includegraphics[width=\linewidth]{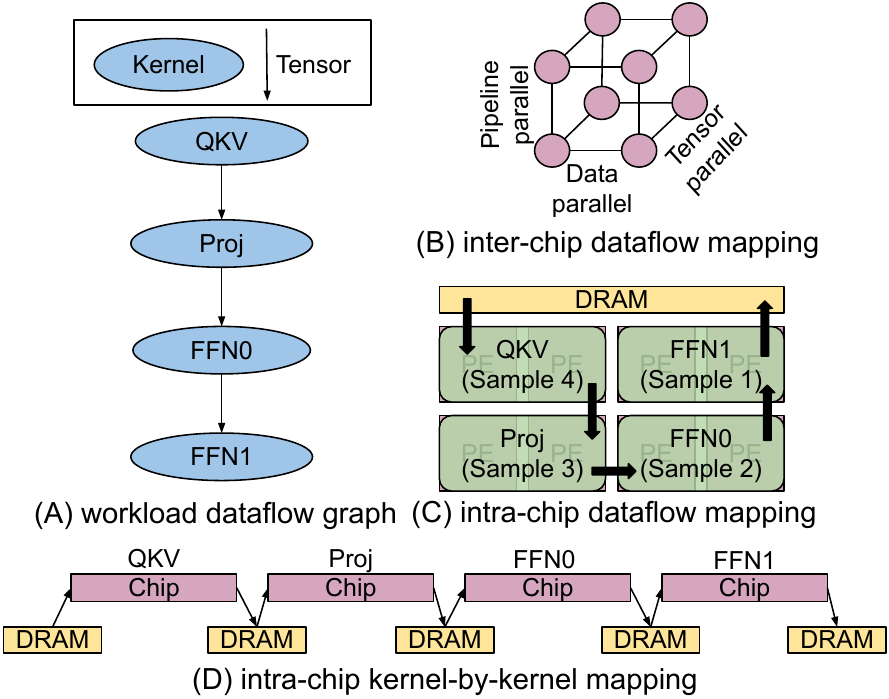}
  \vspace{-10pt}
  \caption{(A) The workload dataflow graph of a single-layer generative pre-training transformer (GPT) model.
  (B) Inter-chip dataflow mapping: parallelization strategies such as tensor parallelism, pipeline parallelism, and data parallelism are used to map a workload onto an eight-chip system.
  (C) Intra-chip dataflow mapping: multiple kernels are fused on-chip and data is pipelined through the kernels in a streaming fashion.
  (D) Intra-chip kernel-by-kernel mapping: kernels are executed sequentially with frequent DRAM accesses between kernels.}
  \label{dataflow}
  \vspace{-10pt}
\end{figure}

To map a workload onto a distributed system, we need to consider that data flows at several levels of memory hierarchies and interconnection networks such as the inter-accelerator (inter-chip) level and single-accelerator (intra-chip) level.
Obtaining an optimal dataflow mapping across accelerators requires different parallelization strategies~\cite{jouppi2023tpu} shown in Figure~\ref{dataflow}B, such as tensor parallelism (TP)~\cite{dean2012large}, pipeline parallelism (PP)~\cite{huang2019gpipe}, and data parallelism (DP)~\cite{zhang1989efficient}.
Additionally, mapping the dataflow graph to each accelerator in the system requires techniques like kernel fusion, tiling, and on-chip pipelining~\cite{10.5555/3600270.3601459, dao2023flashattention2fasterattentionbetter, shah2024flashattention3fastaccurateattention} to achieve high performance, as shown in Figure~\ref{dataflow}C.
To achieve an optimal mapping across the system hierarchies, open questions for research emerge:
\begin{itemize}
    \item What combination of TP/PP/DP degrees leads to the highest performance given a fixed amount of hardware resources?

    \item What is the best way to fuse and tile the kernels on-chip that fits into the hardware resources and simultaneously achieves the highest performance?

    \item How should the inter-chip and intra-chip dataflow mapping balance compute latency with memory and network latency to ensure the system is not bottlenecked by IO?
\end{itemize}

To answer these questions, we propose DFModel---a framework that models and optimizes the performance of any given arbitrary dataflow graph on any given arbitrary system specification.
To model general workloads, DFModel represents a workload as a dataflow graph that describes the different compute kernels and data dependencies in the application. 
Figure~\ref{dataflow}A shows a single layer of a generative pre-training transformer (GPT) model~\cite{radford2019language, brown2020language} represented as a dataflow graph in which vertices represent compute kernels and edges represent tensors.
DFModel then optimizes dataflow mappings at several levels of memory hierarchies and interconnection networks including the inter-accelerator (inter-chip) level and single-accelerator (intra-chip) level.
DFModel formulates the design space as an optimization problem and solves it using the Gurobi optimizer~\cite{gurobi2022gurobi}.
Table~\ref{framework} compares DFModel with performance models from prior works.
Compared to the previous works, DFModel is a more comprehensive model to consider general workloads, optimize dataflow mappings across various hierarchies of systems, and explore the entire system design space.
The contributions of this paper are the following:
\begin{itemize}
\item The representation of general workloads as dataflow graphs and system specifications as constraints.
\item A performance model that encompasses several levels of memory hierarchies and interconnection networks including the inter-chip level and intra-chip level.
\item An automated framework backed by Gurobi~\cite{gurobi2022gurobi} to optimize the mapping of an arbitrary dataflow graph on an arbitrary system specification according to the performance model.
\item DFModel is able to swiftly produce a mapping with provably optimal performance for a trillion-parameter-scale LLM onto a thousand-accelerator datacenter, exploring a design space of size $O(10^{295})$ within 20 minutes on a server with 64 CPUs.
\end{itemize}

We validate DFModel against the measured performance of various DL and HPC workloads on a wide range of industrial systems.
Mappings by DFModel demonstrate an improved average upper bound of 10\% higher performance compared to the measured performance from these systems.
We further validate DFModel against previous performance models such as Calculon~\cite{isaev2023calculon} and Rail-Only~\cite{wang2023optimized} with a 3.1\% to 4.1\% error margin.
The DFModel-optimized dataflow mapping for LLM training achieves $6.13\times$ speedup compared to the non-dataflow mapping from previous performance models and $1.52\times$ speedup compared to a vendor provided dataflow mapping on an industrial system with dataflow architectures~\cite{prabhakar2021sambanova, prabhakar2022sambanova}.
In addition, DFModel shows that dataflow mappings provide a performance upper bound over non-dataflow mappings.

\section{Background}

Mapping complex workloads like LLMs to a distributed system is difficult.
Performance architects need to understand both the workload and the underlying system to optimize the design space and find an optimal mapping.
In this section, we introduce the background required to understand optimizing mappings of one example workload, LLMs, to distributed systems.

\subsection{Large Language Models}
Large language models (LLMs) have demonstrated unprecedented capabilities in a wide range of applications such as bug fixation~\cite{jesse2023large}, education~\cite{10.1145/3643479.3662055}, and drug discovery~\cite{liu2021ai}.
Typical examples of LLMs include GPT2~\cite{radford2019language},  GPT3~\cite{brown2020language}, and LLaMA~\cite{touvron2023llama}.
Each LLM layer consists of an attention layer followed by several linear layers~\cite{brown2020language}.
A high-level dataflow graph for a single-layer GPT model~\cite{radford2019language, brown2020language} is shown in Figure~\ref{dataflow}A.
The attention layer computes the interaction between the query (Q), key (K), and value (V) vectors to compute an attention score.
For example, the commonly used scaled dot-product attention kernel is defined as: $\text{attention}(Q,K,V) = \text{softmax}(QK^\intercal)V$.

\subsection{Accelerator Systems for Training}
Due to the astronomical compute and memory requirements, LLM training is done on a large distributed system of accelerators connected by hierarchical memory and interconnection networks.
Each accelerator is equipped with a large number of floating-point units, on-chip SRAM cache, and off-chip DRAM memory.
Mapping at the inter-chip level requires different parallelization strategies~\cite{jouppi2023tpu} such as TP, PP, and DP, as shown in Figure~\ref{dataflow}B.
Specifically, TP shards the kernels within an LLM layer into multiple accelerators, PP partitions the layers of an LLM into different accelerators, and DP replicates the LLM model multiple times across different accelerators.
Zooming into each accelerator, there are two classes of execution models: kernel-by-kernel and dataflow.
Kernel-by-kernel execution is typically done in instruction-based processor architectures such as NVIDIA's graphics processing unit (GPU)~\cite{choquette2020nvidia, choquette2022nvidia} and Google's tensor processing unit (TPU)~\cite{jouppi2023tpu}.
Kernel-by-kernel execution loads the data from memory to on-chip, executes the kernel, and then stores the results back to memory, as shown in Figure~\ref{dataflow}D.
This incurs more memory traffic and results in compute under-utilization.
Dataflow execution is typically done in spatial architectures like SambaNova's reconfigurable dataflow unit (RDU)~\cite{prabhakar2021sambanova, prabhakar2022sambanova, prabhakar2024sambanova} and Cerebras' wafer-scale engine (WSE)~\cite{lie2021multi, lie2022cerebras}.
Dataflow execution is capable of mapping multiple kernels spatially, fusing them on-chip, and pipelining input data through the kernels in a streaming fashion, as shown in Figure~\ref{dataflow}C.
This spatial computing paradigm results in less memory traffic and improves compute utilization.
Customized mapping like FlashAttention~\cite{10.5555/3600270.3601459, dao2023flashattention2fasterattentionbetter, shah2024flashattention3fastaccurateattention} is needed to map LLM workloads onto dataflow-like execution.

\section{DFModel Methodology}


\subsection{DFModel Overview}
DFModel takes in a system specification and a workload dataflow graph as inputs, as shown in Figure~\ref{DFModel_figure1}.
DFModel partitions the dataflow graph onto different levels of the system hierarchy, where each partition\footnote{In the figures in this paper, par is short for partition.} is a subgraph representing a block of computation.
First, an inter-chip optimization pass divides a dataflow graph into multiple partitions.
The partitions are assigned to multiple chips that run in parallel, and each chip gets one partition of the graph.
Once workloads are partitioned across accelerators, DFModel applies the intra-chip optimization pass to subdivide the partition into even smaller partitions, 
where the smaller partitions run sequentially on each chip.
The final dataflow mapping is the combination of the inter-chip mapping and intra-chip mapping.
For each level of the mapping optimization, DFModel formulates the design space into a mixed-integer programming problem according to a performance model that is fed to Gurobi.
Gurobi~\cite{gurobi2022gurobi} is a constrained optimization solver that takes in a set of user-provided variables, their associated constraints, and an objective function, and produces an optimal assignment of variables to values.
In addition, we chose Gurobi because it is more powerful than academic solvers like ECOS~\cite{Domahidi2013ecos} and POGS~\cite{fougner2018parameter}.

\subsection{DFModel Approaches}

\subsubsection{Formulation Conventions}
We adopt the following mathematical notation when formulating DFModel.
For tensors, matrices are uppercase ($A$), vectors are lowercase with an arrow ($\overrightarrow{a}$), and scalars are lowercase ($a$).
The $\overrightarrow{1}$ denotes a vector of all ones.
Variables are bolded whereas constants are not (e.g. $\mathbf{A}$ versus $A$). 
Booleans, reals, and integers are denoted as: $\mathbb{B}$, $\mathbb{R}$, and $\mathbb{Z}$.
$\mathbf{A}[i, j]$ and $\mathbf{A}[i, :]$ represent indexed access or slicing into a tensor using Numpy-style array notation. 
$\le$, $\ge$, $+$, $-$, $\times$, $/$, $\land$, and $\oplus$ are element-wise binary operations unless otherwise noted.
An operation of a vector by a scalar broadcasts the scalar to every vector element.

\begin{figure*}[t!]
  \centering
  \includegraphics[width=\linewidth]{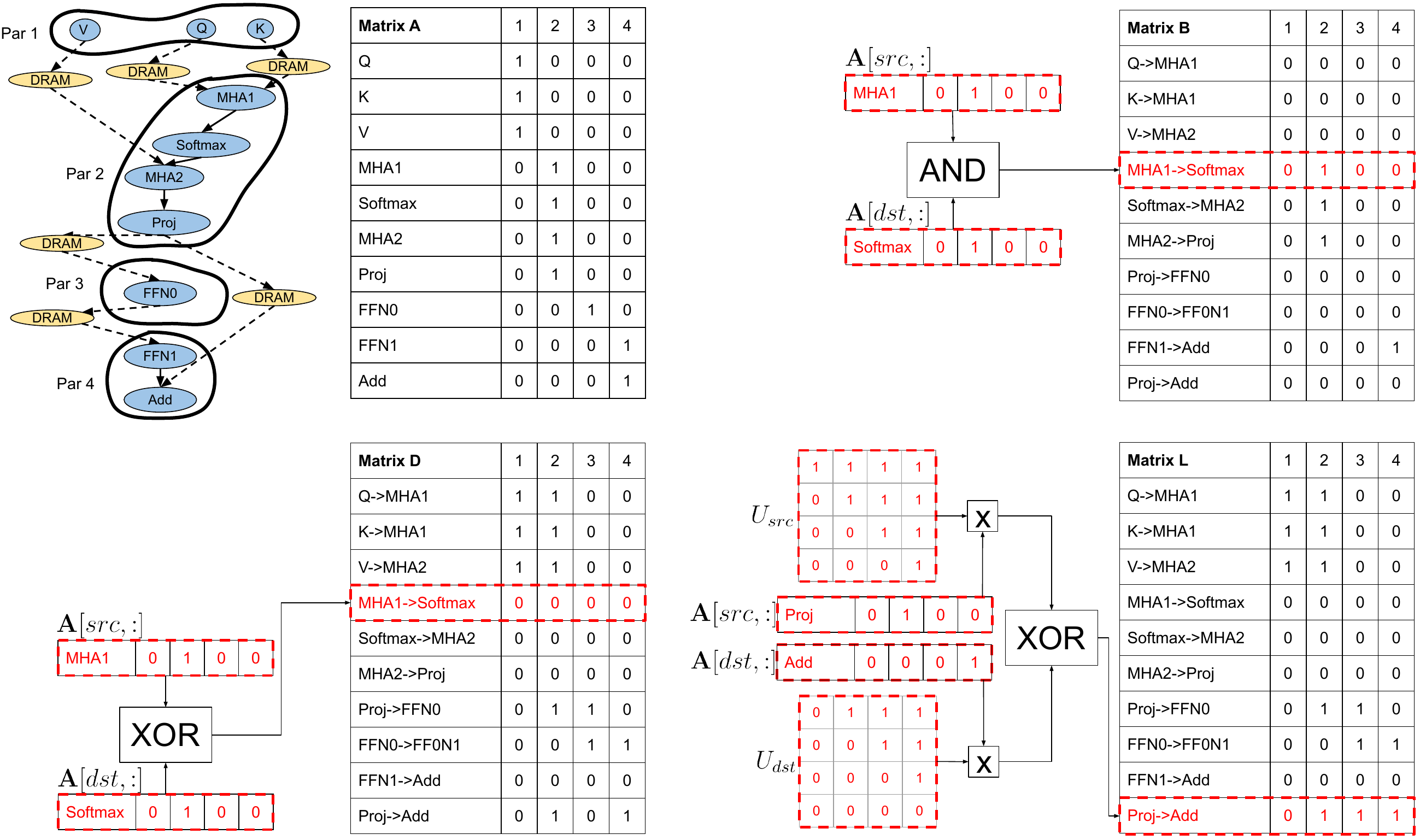}
  \vspace{-10pt}
  \caption{Four assignment matrices used in DFModel.
  Matrix $\mathbf{A}$ encodes the kernel to partition assignment, which is useful for deriving other assignment matrices.
  Matrix $\mathbf{B}$ encodes the tensors which stay within a partition.
  Matrix $\mathbf{D}$ encodes the tensors which cross two different partitions.
  Matrix $\mathbf{L}$ encodes the lifetime of cross-partition tensors.
  Matrix $H$ is not shown due to space constraints.}
  \label{assignment}
  \vspace{-10pt}
\end{figure*}

\begin{table}[t!]
    \centering
    \caption{Constants and variables for inter-chip and intra-chip phases. 
    \\
    \scriptsize{
    *$\mathcal{K}$ and $\mathcal{V}$ are user inputs that describe the kernels and tensors in the workload dataflow graph.
    Derived variables are those whose tensors are dependent on the assignment matrix $\mathbf{A}$ with the purpose of encoding constraints and objectives.}}
    \begin{tabu}{ p{0.7cm}p{5cm}p{0.7cm}p{0.8cm} } 
    
    \rowfont{\bfseries} Name & Description & Type & Def. \\
    \midrule

    $\mathcal{K}$ & Kernels in the dataflow graph & * & Input \\

    $\mathcal{V}$ & Tensors in the dataflow graph & * & Input \\

    $n$ & Number of kernels in the graph & $\mathbb{Z}$ & $|\mathcal{K}|$\\

    $m$ & Number of tensors in the graph & $\mathbb{Z}$ & $|\mathcal{V}|$\\

    $\overrightarrow{f}$ & FLOP in each kernel & $\mathbb{Z}^{n}$ & Input \\
    
    $\overrightarrow{b}$ & Size of each tensor & $\mathbb{R}^{m}$ & Input\\

    $p_{max}$ & Maximum number of partitions to consider & $\mathbb{Z}$ & Input\\

    $\mathbf{A}$ & Kernel to partition assignment & $\mathbb{B}^{n \times p_{\max}}$ & Solver\\

    $\mathbf{B}$ & Intra-partition tensor to partition assignment & $\mathbb{B}^{m \times p_{\max}}$ & Derived\\

    $\mathbf{D}$ & Cross-partition tensor to partition assignment & $\mathbb{B}^{m \times p_{\max}}$ & Derived\\

    \multirow{3}{*}{$\mathbf{L}$} & The lifetime of each tensor, expressed as & \multirow{3}{*}{$\mathbb{B}^{m \times p_{\max}}$} & \multirow{3}{*}{Derived} \\
     & an indicator matrix of whether the tensor & & \\
     & occupies resources in a particular partition & & \\

    \multirow{2}{*}{$\mathbf{H}$} & Indicator matrix encoding the source & \multirow{2}{*}{$\mathbb{B}^{m \times p_{\max}}$} & \multirow{2}{*}{Derived} \\
     & partition of each cross-partition tensor & & \\
    
    \end{tabu}
    \label{graph}
    \vspace{-10pt}
\end{table}

\subsubsection{Common Constants \& Variables}
DFModel uses some common constants and variables in both inter-chip and intra-chip optimization, as shown in Table~\ref{graph}.
The input dataflow graph consists of a directed acyclic graph, with vertices representing computations and edges representing tensors.
In general, the task is to divide this graph into one or more partitions, each representing a block of computation.
In the inter-chip setting, each partition represents a subset of kernels assigned to different accelerators in the systems.
Alternatively in the intra-chip setting, each partition represents a subset of kernels scheduled to execute together in the same accelerator.
The formulation of DFModel optimization centers around the variable matrix $\mathbf{A}$ for which the Gurobi optimizer solves.
Matrix $\mathbf{A}$ encodes the assignment of each kernel in the dataflow graph to the partitions.
We also create other matrices like $\mathbf{B}$, $\mathbf{D}$, $\mathbf{L}$, and $\mathbf{H}$ which are derived from matrix $\mathbf{A}$.
The relationship between the derived matrices and matrix $\mathbf{A}$ are encoded as Gurobi constraints.

\textit{Matrix $\mathbf{A}$.} At the core of our modeling approach is the assignment matrix $\mathbf{A}$ of size [$n$ kernels $\times$ $p_{max}$ partitions].
Each row in $\mathbf{A}$ represents a kernel in the graph, and each column in $\mathbf{A}$ represents a partition.
The vector $\mathbf{A}[i, :]$ is a one-hot vector encoding in which partition kernel $i$ resides.
DFModel requires that each kernel is assigned to exactly one partition, encoded by $\mathbf{A} \overrightarrow{\mathbf{1}} = \overrightarrow{\mathbf{1}}$.
Figure~\ref{assignment} shows an example of a dataflow graph with four partitions and its corresponding matrix $\mathbf{A}$.

\textit{Matrix $\mathbf{B}$.} A tensor is uniquely defined by its upstream kernel and downstream kernel.
A tensor is within a partition if both its upstream kernel and downstream kernel reside in the same partition.
To encode the information for this type of tensor, DFModel uses matrix $\mathbf{B}$ of size [$m$ tensors  $\times p_{max}$ partitions].
The vectors $\mathbf{A}[src,:]$ and $\mathbf{A}[dst,:]$ represent the partition in which a tensor's upstream kernel and downstream kernel reside.
Using the AND operation between the two vectors tells whether the tensor stays within a partition, and if so, in which partition it resides.
Equation~\ref{onchip_tensor} encodes matrix $\mathbf{B}$ and Figure~\ref{assignment} visualizes it. Tensors which stay within a partition are given by: 
\begin{equation} \small
\label{onchip_tensor}
    \forall j = (j_{src} \rightarrow j_{dst}) \in \mathcal{V} : \mathbf{B}[j,:] = \mathbf{A}[src,:] \land \mathbf{A}[dst,:]
\end{equation}

\textit{Matrix $\mathbf{D}$.} A tensor crosses different partitions if its upstream kernel and downstream kernel reside in two different partitions.
To encode cross-partition tensors, DFModel uses matrix $\mathbf{D}$ of size $m \times p_{max}$.
Using the XOR operation between the two vectors tells whether the tensor crosses different partitions, and if so, which two partitions it crosses.
Equation~\ref{offchip_tensor} encodes matrix $\mathbf{D}$, and Figure~\ref{assignment} helps to visualize it. 
\begin{equation} \small
\label{offchip_tensor}
    \forall j = (j_{src} \rightarrow j_{dst}) \in \mathcal{V}: \mathbf{D}[j,:] = \mathbf{A}[src,:] \oplus \mathbf{A}[dst,:]
\end{equation}

\textit{Matrix $\mathbf{L}$.} In addition, DFModel encodes the lifetime of cross-partition tensors in matrix $\mathbf{L}$ since matrix $\mathbf{D}$ only records the start partition and end partition of tensors.
Matrix $\mathbf{L}$ is computed using two upper-triangular constant auxiliary matrices: $U_s[i, j] = i \le j$ and $U_t[i, j] = i < j$.
$\mathbf{A}[src,:] U_{src}$ encodes all ones from the start partition of the tensor to the last partition available and zeros elsewhere.
$\mathbf{A}[dst,:] U_{dst}$ encodes all ones from the end partition of the tensor to the last partition available and zeros elsewhere.
Using the XOR operation between the two vectors selects all partitions between the start and end, which gives the lifetime of a cross-partition tensor.
Note that lifetime only applies to cross-partition tensors, so we need to exclude tensors which stay within a partition.
Equation~\ref{dram_tensor} encodes matrix $\mathbf{L}$ and Figure~\ref{assignment} visualizes it.
\begin{align} \small
\label{dram_tensor}
\begin{split}
\forall j = (j_{src} \rightarrow j_{dst}) \in \mathcal{V}: \\ \mathbf{L}[j,:] = ((\mathbf{A}[src,:] U_{src}) \oplus (\mathbf{A}[dst,:] U_{dst})) \oplus (\mathbf{A}[src,:] \land \mathbf{A}[dst,:])\\
\end{split}
\end{align}

\textit{Matrix $\mathbf{H}$.} DFModel uses matrix $\mathbf{H}$ to encode the tensor placement based on its upstream kernel placement. Matrix $\mathbf{H}$ is formulated in Equation~\ref{h_tensor} and visualized in Figure~\ref{assignment}.
\begin{equation} \small
\label{h_tensor} 
     \forall j = (j_{src} \rightarrow j_{dst}) \in \mathcal{V}: \mathbf{H}[j,:] = \mathbf{A}[src,:]
\end{equation}

\section{Inter-Chip Optimization}

\begin{figure}[t!]
  \centering
  \includegraphics[width=\linewidth]{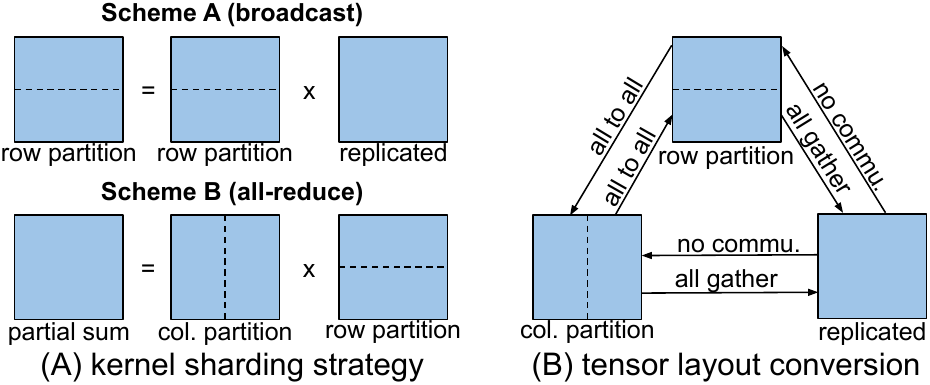}
  \vspace{-10pt}
  \caption{Kernel sharding results in two types of communication cost: communication inherent to the kernel in (A) and tensor layout conversion in (B).
  Using matrix multiplication as an example, two sharding strategies in (A) shard the tensors in the kernel along different dimensions and incur different communication.
  For each tensor, different tensor layout conversions in (B) incur different communication.}
  \vspace{-10pt}
  \label{sharding}
\end{figure}

The inter-chip modeling component of DFModel takes in a program graph alongside a distributed system specification as input.
Using these inputs, the inter-chip optimization pass shards and partitions the dataflow graph into multiple chips.
Accounting for sharding costs, format conversion costs, and network characteristics alongside the computational characteristics of each kernel, DFModel jointly optimizes kernel placement and sharding strategy selection to produce a mapping that maximizes system performance.
Figure~\ref{DFModel_figure2} shows the detailed flow diagram of DFModel.

\subsection{Performance Model}
The inter-chip optimization pass considers the TP and PP parallelization dimensions in a system.
TP shards each kernel in a dataflow graph and PP partitions the dataflow graph into multiple pipeline stages across different chips, where each stage executes a subgraph of kernels.
The PP partitioning introduces point-to-point communication between pipeline stages.

Kernel sharding in TP introduces two types of communication including communication inherent for each kernel and communication for tensor layout conversions.
Figure~\ref{sharding} shows two basic sharding schemes for a matrix-multiplication kernel across two accelerators.
Scheme A in Figure~\ref{sharding} shards rows of the first matrix while replicating the second matrix to both accelerators.
Since the output matrix is naturally sharded by rows, a broadcast communication of the replicated tensor is needed.
Scheme B shards rows of the first matrix and columns of the second matrix simultaneously.
In this case, each accelerator has a partial sum of the full output matrix, and an all-reduce communication of the output matrix is needed.

\begin{figure*}[t!]
  \centering
  \includegraphics[width=\linewidth, height=20cm]{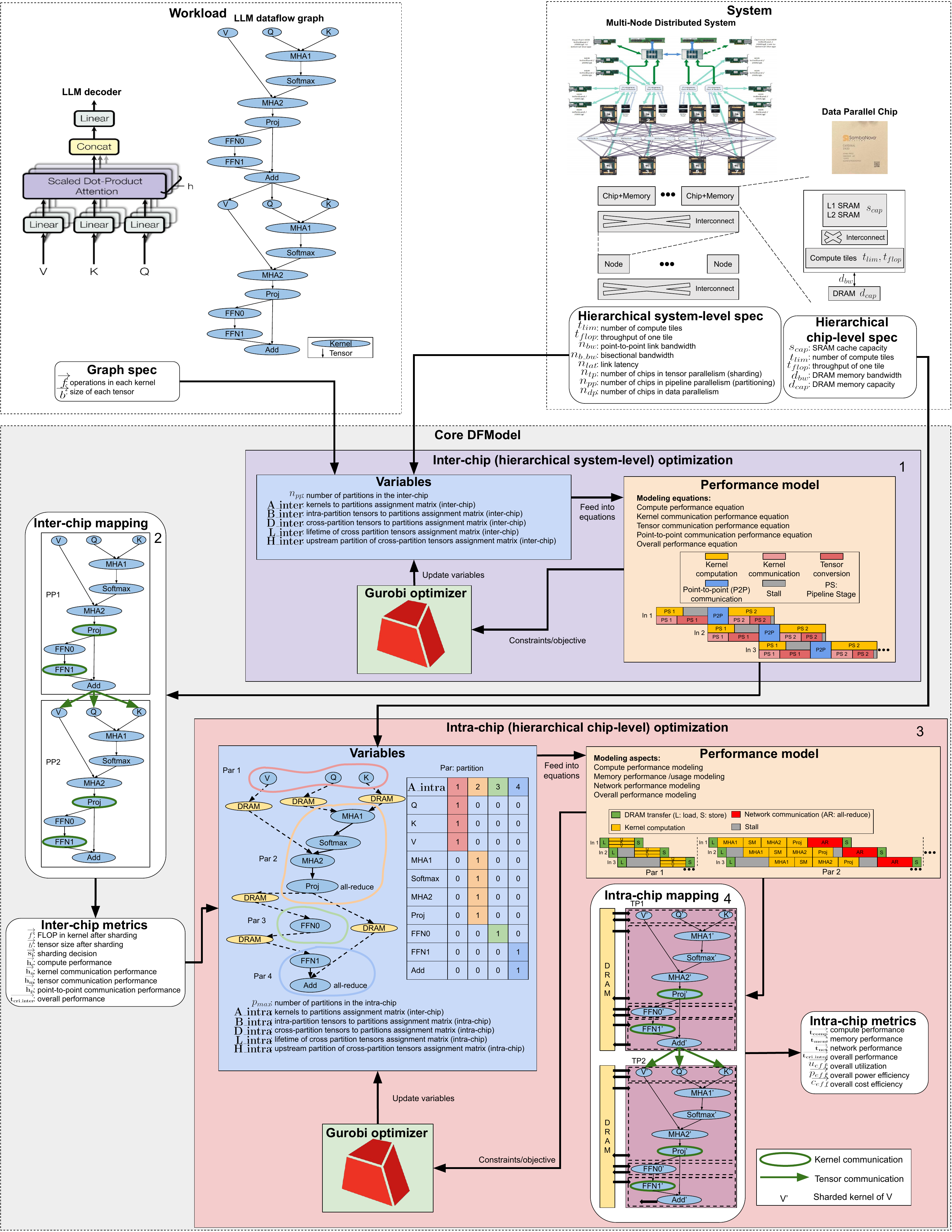}
  \vspace{-10pt}
  \caption{DFModel takes in a workload and a system as inputs.
  The workload is a dataflow graph and the system is a multi-node distributed system composed of several layers of hierarchical memories, interconnection networks, and compute nodes/accelerators.
  Each compute node/accelerator is a data-parallel component with on-chip compute units, hierarchical memories, and a main memory.
  DFModel then undergoes two optimization layers, inter-chip layer (1) and intra-chip layer (3), to find the best dataflow mappings.
  In (1), DF Model optimizes at the hierarchical distributed system level. To do so, DFModel takes the dataflow graph description and multi-node distributed system specification as inputs and combines them with internal assignment variables.
  All the variables are then fed into multiple performance equations modeling various aspects of a distributed system.
  The equations are then encoded as constraints and objectives in Gurobi so that Gurobi iteratively updates the variables.
  This process is repeated continuously until the objective is reached.
  Then the inter-chip mapping and its associated variables (2) are fed to the intra-chip optimization level (3).
  In (3), the inputs are combined with the specification of a data-parallel chip, and (3) iteratively solves the problem similar to the inter-chip level.
  Eventually, the inter-chip level produces the final dataflow mapping (4).
  }
  \label{DFModel_figure2}
  \vspace{-10pt}
\end{figure*}

In DFModel's performance model, kernel computation time can be fully overlapped with kernel/tensor communication time for a given input, as shown in Figure~\ref{DFModel_figure2}.
As a result, DFModel's performance model needs to calculate pipeline stages and point-to-point communication time, which can be overlapped with each other for different inputs as shown in Figure~\ref{DFModel_figure2}.


\subsection{Formulation}

DFModel quantifies the compute and network time in its formulation by using the constants and variables shown in Table~\ref{table_interchip_opt}.
First, DFModel ensures that each kernel only has one chosen sharding strategy.
$\overrightarrow{\mathbf{s_i}}$ is a one-hot vector encoding the sharding strategy of kernel $i$.
Multiplying this vector with a vector of ones should yield the value one to guarantee the unique sharding of a kernel, denoted by $ \forall i \in [1, n]: \overrightarrow{\mathbf{s_i}} \overrightarrow{\mathbf{1}} = 1$.

\begin{table}[t!]
    \centering
    \caption{Some constants and variables for the inter-chip optimization.
    \scriptsize{
    *For a kernel $k_i$ with $z_i$ sharding schemes, $\overrightarrow{c_{i}} \in \mathbb{R}^{z_i}$ contains the pre-computed communication cost of evaluating the operation with each scheme. $\mathbf{s_i} \in \mathbb{B}^{z_i}$ is a one-hot assignment vector denoting which of the $z$ schemes is selected.
    \\
    **For a tensor $v$ produced by kernel $j_{src}$ and consumed by kernel $j_{dst}$, the layout transition cost $C_{v} \in \mathbb{R}^{z_{src} \times z_{dst}}$ contains the pre-computed costs of all possible layout transformations.}}

    \begin{tabu}{ p{0.7cm}p{6cm}p{0.7cm} } 
    
    \rowfont{\bfseries} Name & Description & Type \\
    \midrule

    $t_{lim}$ & Compute tile limit & $\mathbb{Z}$\\

    $t_{flop}$ & Compute throughput of one tile & $\mathbb{R}$\\
    
    $s_{cap}$ & SRAM cache capacity limit & $\mathbb{R}$\\
    
    $d_{cap}$ & DRAM memory capacity limit & $\mathbb{R}$\\
    
    $d_{bw}$ & DRAM memory bandwidth & $\mathbb{R}$\\

    $n_{bw}$ & Network bandwidth & $\mathbb{R}$\\

    $n_{tp}$ & Number of chips in tensor parallelism (sharding) & $\mathbb{Z}$\\

    $n_{pp}$ & Number of chips in pipeline parallelism (partitioning) & $\mathbb{Z}$\\

    $n_{dp}$ & Number of chips in data parallelism & $\mathbb{Z}$\\

    $\overrightarrow{\mathbf{s_i}}$ & The sharding assignment vector for kernel $k_i$ & * \\

    $\overrightarrow{c_i}$ & Inherent cost of each sharding strategy for kernel $k_i$ & *\\

    \multirow{3}{*}{$C_{j}$} & A collection of matrices containing transition costs & \multirow{3}{*}{**}\\
    & for the $j^{th}$ tensor, based on the producer and consumer &\\
    & sharding schemes &\\

    $\overrightarrow{\mathbf{h_{c}}}$ & Computation time of each kernel & $\mathbb{R}^{n}$ \\

    $\overrightarrow{\mathbf{h_{n}}}$ & Communication time of each kernel & $\mathbb{R}^{n}$ \\

    $\overrightarrow{\mathbf{h_{m}}}$ & Communication time of each tensor & $\mathbb{R}^{m}$ \\

    $\overrightarrow{\mathbf{h_{p}}}$ & Point-to-point time of each tensor & $\mathbb{R}^{m}$ \\

    $\overrightarrow{\mathbf{t_{cri\_inter}}}$ & Per-partition critical time in the inter-chip & $\mathbb{R}^{n_{pp}}$\\

    \end{tabu}
    \label{table_interchip_opt}
    \vspace{-10pt}
\end{table}

\subsubsection{Compute Modeling}
DFModel then calculates the per-kernel computation time.
$\overrightarrow{f}[i]$ represents the amount of floating-point operations (FLOP) in a kernel $i$. A specified system is capable of $n_{tp} \times t_{lim} \times t_{flop}$  floating-point operation per second (FLOPS) where  $n_{tp}$ represents the number of chips involved, each having $t_{lim}$ compute tiles with a throughput of $t_{flop}$ per tile.
Dividing FLOP by the actual FLOPS gives a performance upper bound, denoted by $\forall i \in [1, n]: \overrightarrow{\mathbf{h_{c}}}[i] = \frac{\overrightarrow{f}[i]}{n_{tp} \times t_{lim} \times t_{flop}}$.

\subsubsection{Communication Modeling}
For communication costs, DFModel uses adaptations of network models for collective communication from prior work~\cite{thakur2005optimization, cho2019blueconnect}. DFModel populates the vector $\overrightarrow{c_{i}}$ with the inherent communication cost of different sharding strategies for kernel $i$ on a user-defined interconnection topology.
DFModel also populates the matrix $C_{j}$ with the communication cost of layout conversions for tensor $j$. The upstream kernel of the tensor $j$ chooses a sharding strategy $\overrightarrow{\mathbf{{s}_{j_{src}}}}$ and the downstream kernel of that tensor chooses a sharding strategy $\overrightarrow{\mathbf{{s}_{j_{dst}}}}$.
This formulation allows DFModel to consider \emph{any} set of execution strategies on arbitrarily composed networks for which these parameters are known.
Equations~\ref{kernel_commu} and~\ref{tensor_commu} show the formulation of kernel and tensor communication latency respectively, where the $lookup$ operation performs a one-hot lookup which selects an element from an array (such as $\vec{c_i}$) according to the position of the one in the query vector (such as $\overrightarrow{\mathbf{s_i}}$).
\begin{equation} \small
\label{kernel_commu}
\forall i \in [1, n]: \overrightarrow{\mathbf{h_{n}}}[i] = lookup(\vec{c_i}, \overrightarrow{\mathbf{s_i}})
\end{equation}
\begin{equation} \small
\label{tensor_commu}
\forall j = (j_{src} \rightarrow j_{dst}) \in \mathcal{V}: \overrightarrow{\mathbf{h_{m}}}[j] = lookup(C_j, \overrightarrow{\mathbf{{s}_{j_{dst}}}}, \overrightarrow{\mathbf{{s}_{j_{src}}}})
\end{equation}

Point-to-point communication happens when a tensor crosses two different partitions.
Point-to-point communication latency can be modeled by dividing the size of tensor $j$, $\overrightarrow{b}[j]$, by the network bandwidth, $n_{bw}$, according to~\cite{isaev2023calculon} and formulated as $\forall j \in [1, m]: \overrightarrow{\mathbf{h_{p}}}[k] = \frac{\overrightarrow{b}[j]}{n_{bw}}$.

DFModel follows the performance model in Figure~\ref{DFModel_figure2} to calculate the latency of each partition in PP.
For partition $i$, $\overrightarrow{\mathbf{t_{comp}}} = (\mathbf{A}^\intercal \overrightarrow{\mathbf{h_{c}}})[i]$ gives the kernel computation latency; $\overrightarrow{\mathbf{t_{net\_k}}} =(\mathbf{A}^\intercal \overrightarrow{\mathbf{h_{n}}})[i]$ gives the kernel communication latency; $\overrightarrow{\mathbf{t_{net\_t}}} =(\mathbf{H}^\intercal \overrightarrow{\mathbf{h_{m}}})[i]$ gives the tensor communication latency; $\overrightarrow{\mathbf{t_{p2p}}} = (\mathbf{L}^\intercal \overrightarrow{\mathbf{h_{p}}})[i]$ gives the point-to-point communication latency since point-to-point communication happens across all partitions between the start and end partition.
The total network latency for the chips within a partition is denoted as the sum of the kernel and tensor communication latency for each partition $\overrightarrow{\mathbf{t_{net}}} = \overrightarrow{\mathbf{t_{net\_k}}} + \overrightarrow{\mathbf{t_{net\_t}}}$.
Thus, according to the performance model in Figure~\ref{DFModel_figure2}, the critical time bottlenecking a partition is:
\begin{equation} \small
\label{pipe_stage}
\forall i \in [1, n_{pp}]: \overrightarrow{\mathbf{t_{cri\_inter}}}[i] = max(\overrightarrow{\mathbf{t_{comp}}}[i], \overrightarrow{\mathbf{t_{net}}}[i], \overrightarrow{\mathbf{t_{p2p}}}[i])
\end{equation}

The final objective of DFModel's inter-chip optimization is to minimize $\mathbf{t_{cri\_inter}}$ the critical partition of each partition, formulated as $\text{minimize}: \text{max}_{i \in [1, n_{pp}]} \overrightarrow{\mathbf{t_{cri\_inter}}}[i]$

\subsection{DFModel Assumptions}
DFModel assumes all kernels in the graph are throughput-oriented dense linear algebra kernels and the pipeline in each partition reaches steady-state.
Tensors are assumed to have a single producer (upstream) and a single consumer (downstream), and tensors that are used by multiple consumers are replicated.
For off-chip components, DFModel assumes DRAM and network bandwidth can be fully utilized.
For the off-chip interconnection topology, DFModel borrows the compositional network approach from ASTRA-sim~\cite{rashidi2020astra}: a multidimensional interconnection topology is defined by hierarchically composing multiple one-dimensional topologies such as ring, fully-connected, and switch.
DFModel assumes that one network dimension can only be assigned to one parallelization strategy so subdividing a network dimension is not allowed.

\section{Intra-Chip Optimization}
Once the inter-chip optimization pass of DFModel is complete, every chip will have an assigned subset of computation. Then, DFModel will perform its intra-chip optimization pass, which further subdivides the computation within each chip into multiple partitions that execute on that chip sequentially.
The formulation searches for the partitioning strategy with the highest performance by maximizing the overlap time between compute, memory, and network while satisfying all chip-level constraints.
Figure~\ref{DFModel_figure2} shows the detailed flow diagram of DFModel.

\subsection{Performance Model}
Within each partition, DFModel assigns some on-chip compute resources to each kernel, and the kernels are fully pipelined on-chip between inputs, as shown in Figure~\ref{DFModel_figure2}.
Besides, kernel computation, DRAM transfer, and network communication are also considered in DFModel.
DFModel assumes the DRAM and network time can be fully overlapped with computation time within a partition and the longest time among compute, memory, and network will be the performance bottleneck, as shown in Figure~\ref{DFModel_figure2}.


\begin{table}[b!]
    \centering
    \vspace{-10pt}
    \caption{Variables for the intra-chip optimization pass.}
    \begin{tabu}{ p{1cm}p{5cm}p{1cm} } 
    
    \rowfont{\bfseries} Name & Description & Type \\
    \midrule

    $p_{max}$ & Maximum number of partitions to consider & $\mathbb{Z}$ \\

    $\overrightarrow{f'}$ & FLOP in each kernel after sharding & $\mathbb{Z}^{n}$ \\
    
    $\overrightarrow{b'}$ & Size of each tensor after sharding & $\mathbb{R}^{m}$ \\
    
    $\overrightarrow{\mathbf{u_{c}}}$ & Compute utilization of each kernel & $\mathbb{R}^{n}$ \\
     
    $\overrightarrow{\mathbf{t_{used}}}$ & Per-kernel compute tile usage & $\mathbb{Z}^{n}$\\
    
    $\overrightarrow{\mathbf{t_{comp}}}$ & Per-partition compute time & $\mathbb{R}^{p_{max}}$\\

    $\overrightarrow{\mathbf{t_{mem}}}$ & Per-partition memory time & $\mathbb{R}^{p_{max}}$\\

    $\overrightarrow{\mathbf{t_{net}}}$ & Per-partition network time & $\mathbb{R}^{p_{max}}$\\

    $\overrightarrow{\mathbf{t_{cri\_intra}}}$ & Per-partition critical time in the intra-chip & $\mathbb{R}^{p_{max}}$\\
    
    \end{tabu}
    \label{table_onchip_opt}
\end{table}

\subsection{Formulation}
DFModel uses the variables in Table~\ref{table_onchip_opt} to formulate the compute, memory, and network modeling of a single chip into an optimization problem.

\subsubsection{Compute Modeling}
DFModel assigns some compute tiles to each kernel within a partition, denoted as $\mathbf{t_{used}}$.
$\mathbf{A}^\intercal \overrightarrow{\mathbf{t_{used}}}$ calculates the total number of tiles used in each partition and it should be within the tile limit, denoted as $t_{lim}$.
Therefore, the on-chip compute resource constraint is encoded as $\mathbf{A}^\intercal\overrightarrow{\mathbf{t_{used}}} \le \overrightarrow{t_{lim}}$.
$\mathbf{A}[:, i]$ indicates the kernels in partition $i$, and $\overrightarrow{f'}$ indicates the FLOP in each kernel.
The term $\mathbf{A}[:, i] \times \overrightarrow{f'}$ represents the FLOP per kernel in partition $i$.
In addition, DFModel calculates a compute utilization factor $\overrightarrow{\mathbf{u_{c}}}$ which denotes the utilization of the kernel on the given compute resources, following empirical performance equations for compute cycles from~\cite{samajdar2020systematic}.
Then the term $\overrightarrow{\mathbf{t_{used}}} \times t_{flop} \times \overrightarrow{\mathbf{u_{c}}}$ represents the hardware compute resources in FLOPS assigned to each kernel.
Dividing the two terms and taking the maximum value give the critical kernel latency for each partition, formulated as $\forall_{i \in [1, p_{max}]}: \overrightarrow{\mathbf{t_{comp}}}[i] = max(\frac{\mathbf{A}[:, i] \times \overrightarrow{f'}}{\overrightarrow{\mathbf{t_{used}}} \times t_{flop} \times \overrightarrow{\mathbf{u_{c}}}})$.

\subsubsection{Memory Modeling}
For memory modeling, DFModel considers on-chip SRAM capacity usage, off-chip DRAM usage, and DRAM transfer time for all partitions.
For on-chip SRAM modeling, matrix $\mathbf{B}$ is helpful since it records tensors whose upstream kernel and downstream kernel stay within the same partition, indicating that the tensor stays on-chip.
With $\overrightarrow{b'}$ indicating the size of each tensor, $\mathbf{B}^\intercal \overrightarrow{b'}$ calculates the on-chip SRAM usage for all partitions, which should be within the SRAM capacity limit $s_{cap}$.
The on-chip SRAM capacity constraint is therefore encoded as $\mathbf{B}^\intercal \overrightarrow{b'} \le \overrightarrow{s_{cap}}$.

The off-chip DRAM capacity constraint is denoted as $\mathbf{L}^\intercal \overrightarrow{b'} \le \overrightarrow{d_{cap}}$.
Matrix L records the lifetimes of the tensors and is helpful for modeling off-chip DRAM usage since DRAM must contain the tensors during their lifetimes. The term $\mathbf{L}^\intercal \overrightarrow{b'}$ represents the DRAM usage for all partitions, which must be within the DRAM capacity limit $d_{cap}$.

To model the DRAM transfer time per partition, matrix $\mathbf{D}$ is helpful since it records the tensors which cross two different partitions.
The tensor needs to be stored to DRAM by the upstream partition and loaded from DRAM by the downstream partition, both of which contribute to DRAM transfer latency.
The term $\mathbf{D}^\intercal \overrightarrow{b'}$ represents the DRAM transfer size and dividing it by $d_{bw}$ calculates the DRAM transfer latency for all partitions, formulated as $\overrightarrow{\mathbf{t_{mem}}} = \frac{\mathbf{D}^\intercal \overrightarrow{b'}}{\overrightarrow{d_{bw}}}$.

\subsubsection{Network Modeling}
Inherited from Equations~\ref{kernel_commu} and~\ref{tensor_commu}, DFModel computes the network communication latency per partition using the previously calculated network parameters $\overrightarrow{c_i}, C_j$ and sharding decision vector $\overrightarrow{\mathbf{s_i}}$.
The intra-chip network modeling also uses the network latency term $\overrightarrow{\mathbf{t_{net}}}$ from Equation~\ref{pipe_stage}.

\subsubsection{Overall Performance Modeling}
After figuring out the per-partition compute, memory, and network latency, we are able to calculate the overall throughput of the pipeline.
We follow the performance model shown in Figure~\ref{DFModel_figure2}.
For every partition, DRAM transfer latency, kernel computation latency, and network communication latency can be fully pipelined.
The critical latency per partition is the maximum of the three terms: $\overrightarrow{\mathbf{t_{cri\_intra}}} = max(\overrightarrow{\mathbf{t_{comp}}}, \overrightarrow{\mathbf{t_{mem}}}, \overrightarrow{\mathbf{t_{net}}})$.
DFModel's intra-chip optimization pass tries to minimize the sum of critical latency across all partitions, formulated as $\text{minimize}: \sum_{i \in [1, p_{max}]} \overrightarrow{\mathbf{t_{cri\_intra}}}[i]$.

\section{DFModel Evaluation}

DFModel is designed to model a wide range of distributed systems with different accelerator architectures including NVIDIA GPUs~\cite{choquette2020nvidia, choquette2022nvidia}, Google TPUs~\cite{jouppi2023tpu}, SambaNova RDUs~\cite{prabhakar2021sambanova, prabhakar2022sambanova}, and Cerebras WSEs~\cite{lie2021multi, lie2022cerebras}.
The architectural parameters of each accelerator are listed in Table~\ref{chips}.
In this section, we first validate DFModel's accuracy by comparing it against prior performance models~\cite{isaev2023calculon, wang2023optimized} and the measured performance from real systems.
We then provide a complete design space exploration (DSE) for four workloads on various system setups.

\begin{table}[b!]
    \centering
    \vspace{-10pt}
    \caption{Four different accelerator chips.}
    \begin{tabu}{ p{2.8cm}p{2cm}p{2cm}}

    \rowfont{\bfseries} Accelerator & Compute Thru. & SRAM Capacity \\
    \midrule

    NVIDIA H100 GPU & 993 TFLOPS & 113 MB \\
     
    Google TPU v4 & 275 TFLOPS & 160 MB \\
   
    SambaNova SN30 RDU & 614 TFLOPS & 640 MB\\

    Cerebras WSE-2 & 7500 TFLOPS & 40 GB\\
    
    \end{tabu}
    \label{chips}
\end{table}

\begin{figure*}[t!]
  \centering
  \includegraphics[width=\linewidth]{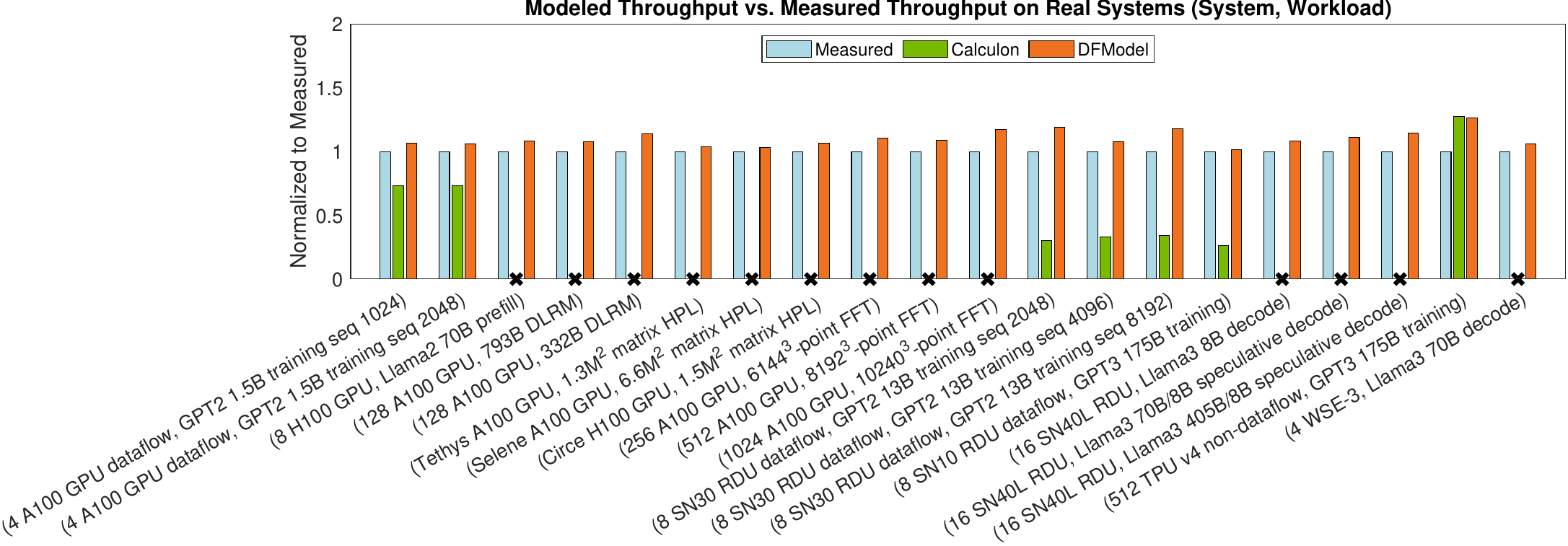}
  \vspace{-10pt}
  \caption{Comparison of the modeled performance from DFModel and Calculon of various workloads on different accelerator systems against the measured performance from industrial systems.
  DFModel performance is 10\% higher than the measured performance.
  Calculon performance for dataflow execution is 60\% lower than the measured performance.}
  \label{overall_comparison}
  \vspace{-10pt}
\end{figure*}

\begin{figure}[t!]
  \centering
  \includegraphics[width=\linewidth]{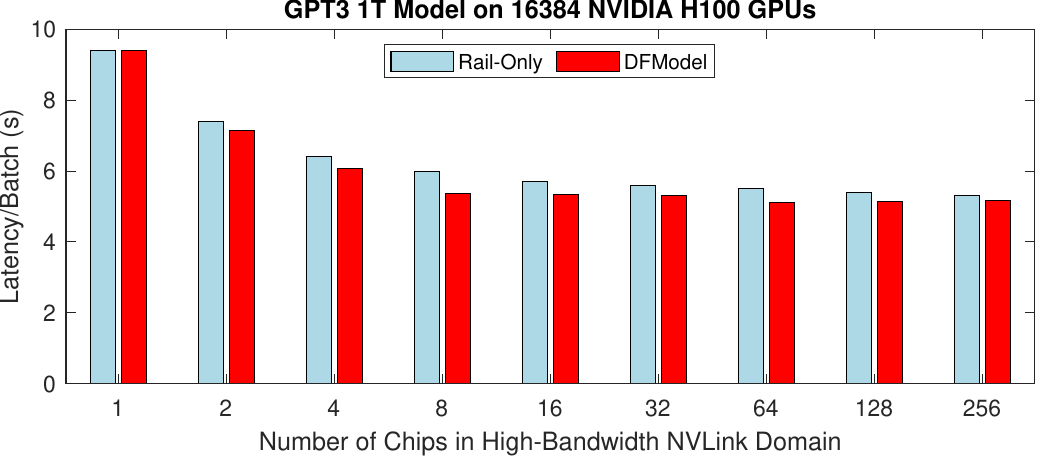}
  \vspace{-10pt}
  \caption{Comparison of DFModel-estimated performance against Rail-Only-estimated performance~\cite{wang2023optimized}.
  We fix the total number of NVIDIA H100 GPUs and sweep the number of GPUs in the high-bandwidth NVLink domain.
  DFModel-estimated performance averages a 3.1\% error margin compared to Rail-Only-estimated performance.}
  \label{rail_only}
  \vspace{-10pt}
\end{figure}

\begin{figure}[t!]
  \centering
  \includegraphics[width=\linewidth]{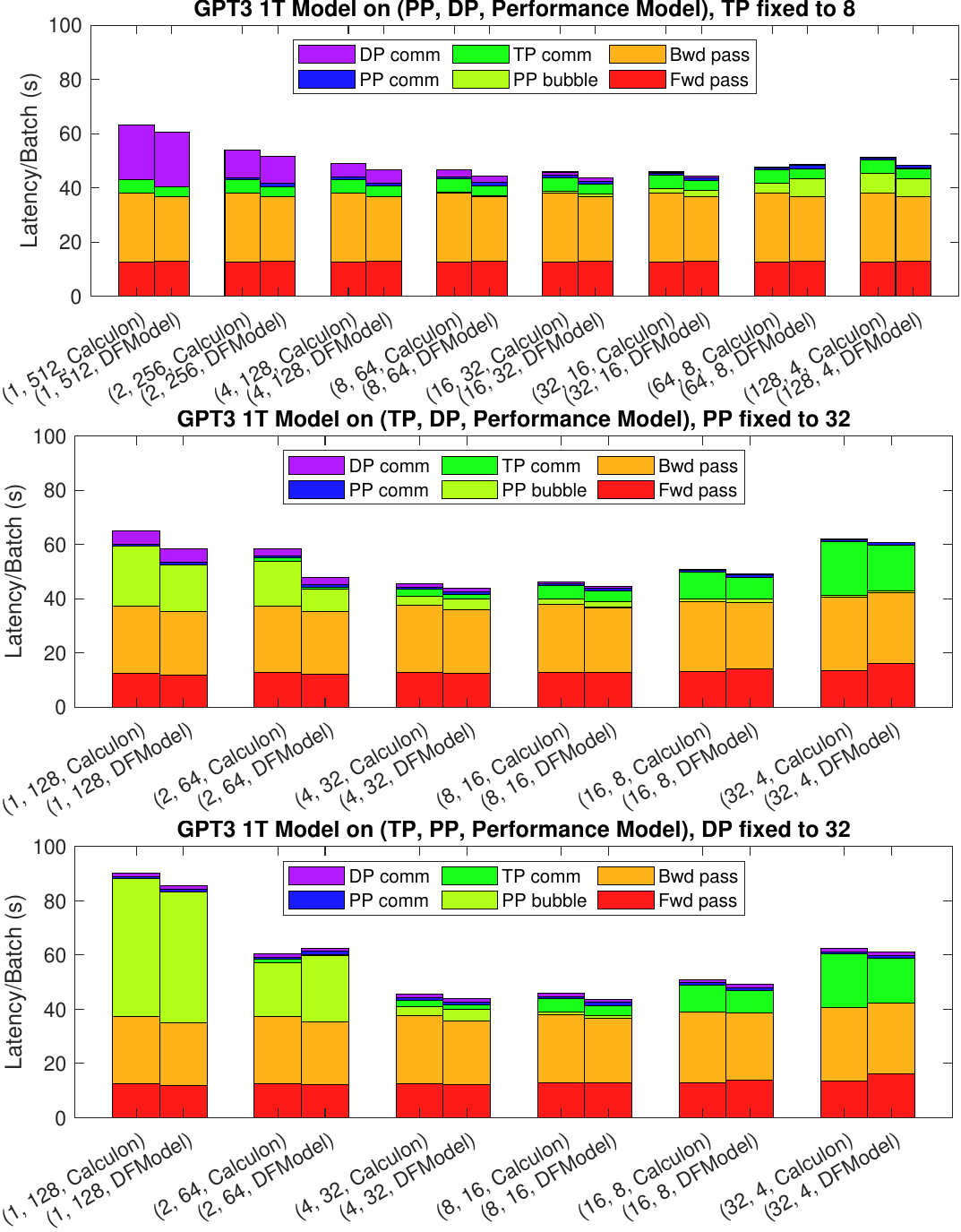}
  \vspace{-10pt}
  \caption{Comparison of DFModel-estimated performance against Calculon-estimated performance~\cite{isaev2023calculon}.
  We fix the total number of NVIDIA A100 GPUs and sweep the number of chips used for different inter-chip parallelization strategies.
  DFModel-estimated performance averages a 4.1\% error margin compared to Calculon-estimated performance.}
  \label{calculon_three}
  \vspace{-10pt}
\end{figure}

\begin{figure}[t!]
    \centering
    \includegraphics[width=\linewidth]{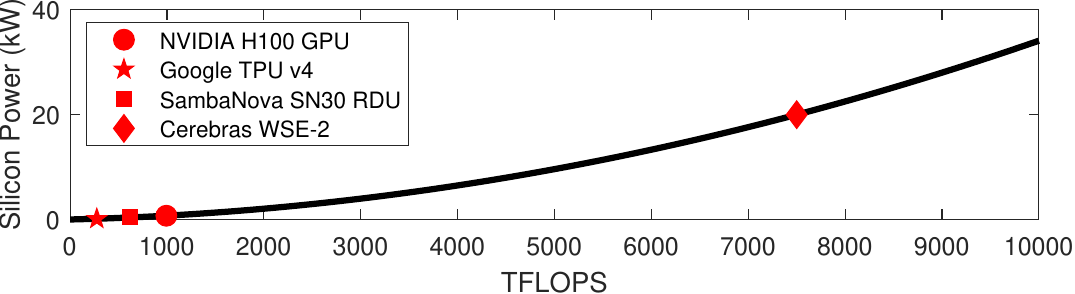}
    \vspace{-10pt}
    \caption{Silicon power vs. compute throughput for four different accelerators.
    A polynomial regression curve connects the data points ($Power: Y = 3 \times 10^{-7}X^{2} - 4.3 \times 10^{-4}X + 0.04$). 
    The curve shows that there is a superlinear relationship between compute throughput and silicon power.
    Based on our analysis, the trend is similar for silicon price, which we do not show due to space constraints.}
    \label{cost_function}
    \vspace{-10pt}
\end{figure}

\subsection{Validation of DFModel Against Previous Models}
For LLM workloads, DFModel's inter-chip optimization suggests that the sharding strategy with the lowest communication cost requires four all-reduces per iteration (forward and backward) per layer, which is confirmed by expert hand partitioning from prior work~\cite{shoeybi2019megatron, narayanan2021efficient}.
We further compared DFModel-estimated performance against two performance models Calculon~\cite{isaev2023calculon} and Rail-Only~\cite{wang2023optimized}.
Since previous performance models only support the kernel-by-kernel non-dataflow mapping (shown in Figure~\ref{dataflow}C), 
we configured DFModel to model non-dataflow mappings for a fair comparison.
In Figure~\ref{calculon_three}, we fix the total number of NVIDIA A100 GPUs in the system and sweep the number of chips used for different inter-chip parallelization strategies.
The figure shows the comparison of DFModel-estimated performance against Calculon-estimated performance for the GPT3 1T model.
The figure gives a detailed latency breakdown between forward pass, backward pass, pipeline bubble, and the communication of different parallelization strategies.
On average, DFModel-estimated performance has a 4.1\% error margin when compared to Calculon-estimated performance.
Rail-Only~\cite{wang2023optimized} proposes a network design with reduced interconnection links that do not compromise LLM training performance.
In Figure~\ref{rail_only}, we fix the total number of NVIDIA H100 GPUs in the system and sweep the number of GPUs in the high-bandwidth NVLink domain.
The figure shows the comparison of DFModel-estimated performance against Rail-Only-estimated performance for the GPT3 1T model.
DFModel-estimated performance averages a 3.1\% error margin.

\subsection{Validation of DFModel Against Measured Performance}
Figure~\ref{overall_comparison} shows the comparison of the modeled performance of four different workloads (LLM/DLRM/HPL/FFT) on different accelerator systems against measured performance from various industrial systems.
We gather the LLM performance from~\cite{emani2023comprehensive, jouppi2023tpu, prabhakar2024sambanova, lie2024wafer, mlperf2024inference}, DLRM performance from from~\cite{10.1145/3470496.3533727}, HPL performance from~\cite{Selene, Circe, Tethys}, and FFT performance from~\cite{cufftmp}.
The mapping by DFModel demonstrates an improved average upper bound of 10\% higher performance when compared to the measured performance from these systems.
In general, DFModel predicts higher performance than the measured performance on real systems because real hardware is complicated, and certain system-level overheads such as driver latency, kernel launching, optimizer steps, logging, and checkpointing~\cite{narayanan2021efficient} cannot be captured by a first-order analytical model.
In addition, we plot the modeled performance from Calculon~\cite{isaev2023calculon}.
Compared to DFModel, Calculon~\cite{isaev2023calculon} has several limitations:
\begin{itemize}
    \item Calculon only models LLM workloads, not DLRM/HPL/FFT workloads.
    \item Calculon only models kernel-by-kernel execution, not dataflow execution.
    For dataflow execution of LLMs, Calculon predicts 60\% lower performance compared to the measured performance.
\end{itemize}

\begin{figure*}[t!]
    \centering
    \includegraphics[width=\linewidth]{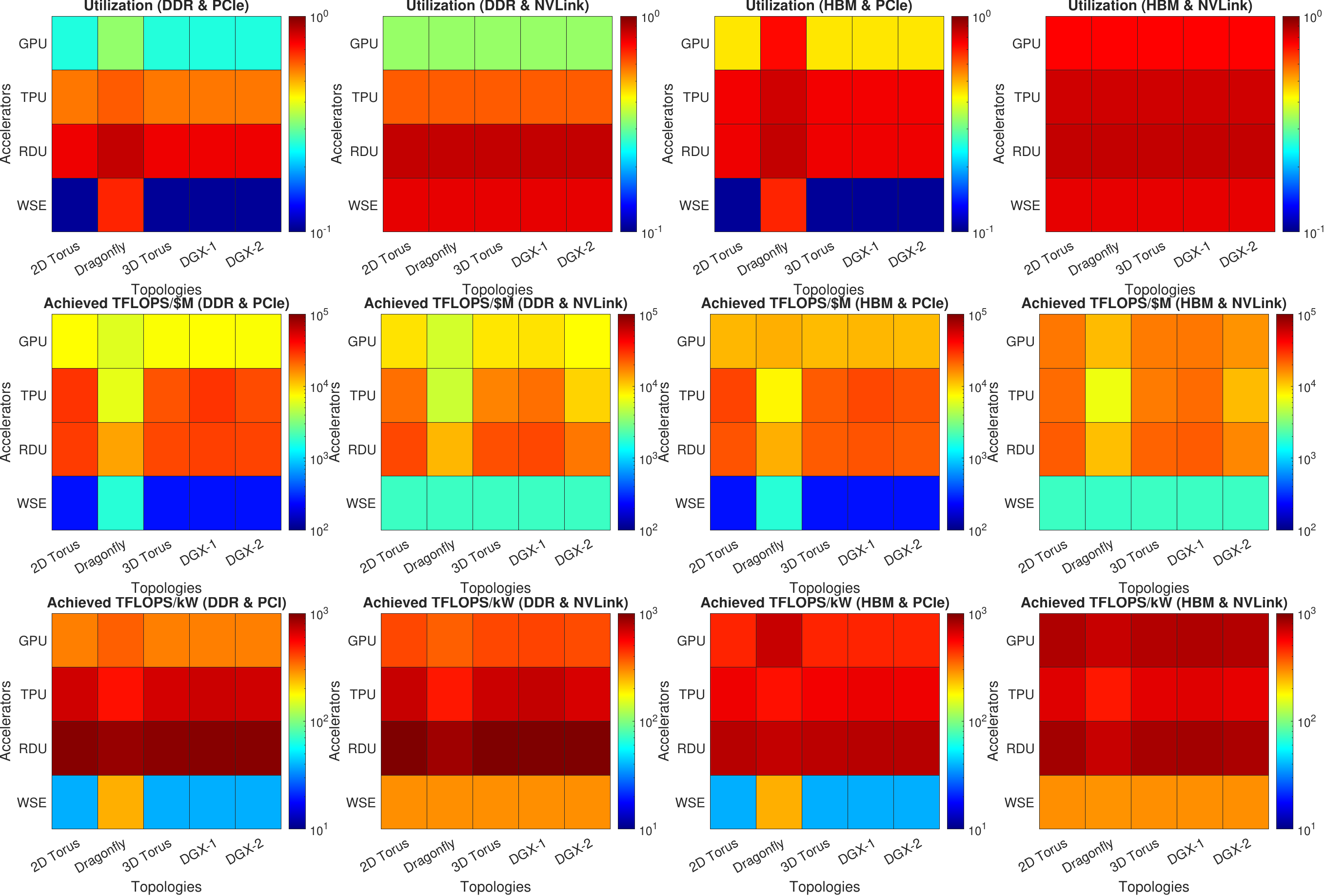}
    \vspace{-10pt}
    \caption{The heat map shows utilization, power efficiency, and cost efficiency of GPT3 1T on 1024 accelerators with different interconnection topologies and memory/network technologies.}
    \label{LLM_heatmap}
    \vspace{-10pt}
\end{figure*}

\begin{figure*}[t!]
    \centering
    \includegraphics[width=\linewidth]{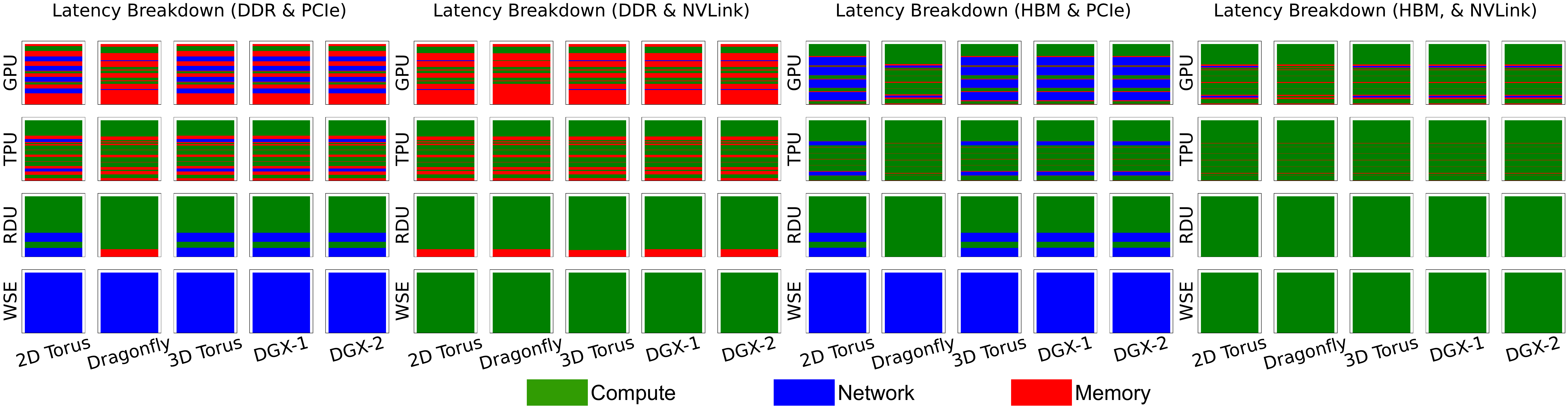}
    \vspace{-10pt}
    \caption{The plot shows the breakdown of compute/memory/network latency of GPT3 1T on 1024 accelerators with different interconnection topologies and memory/network technologies.}
    \label{profiler_LLM}]
    \vspace{-10pt}
\end{figure*}

\begin{figure*}[t!]
    \centering
    \includegraphics[width=\linewidth]{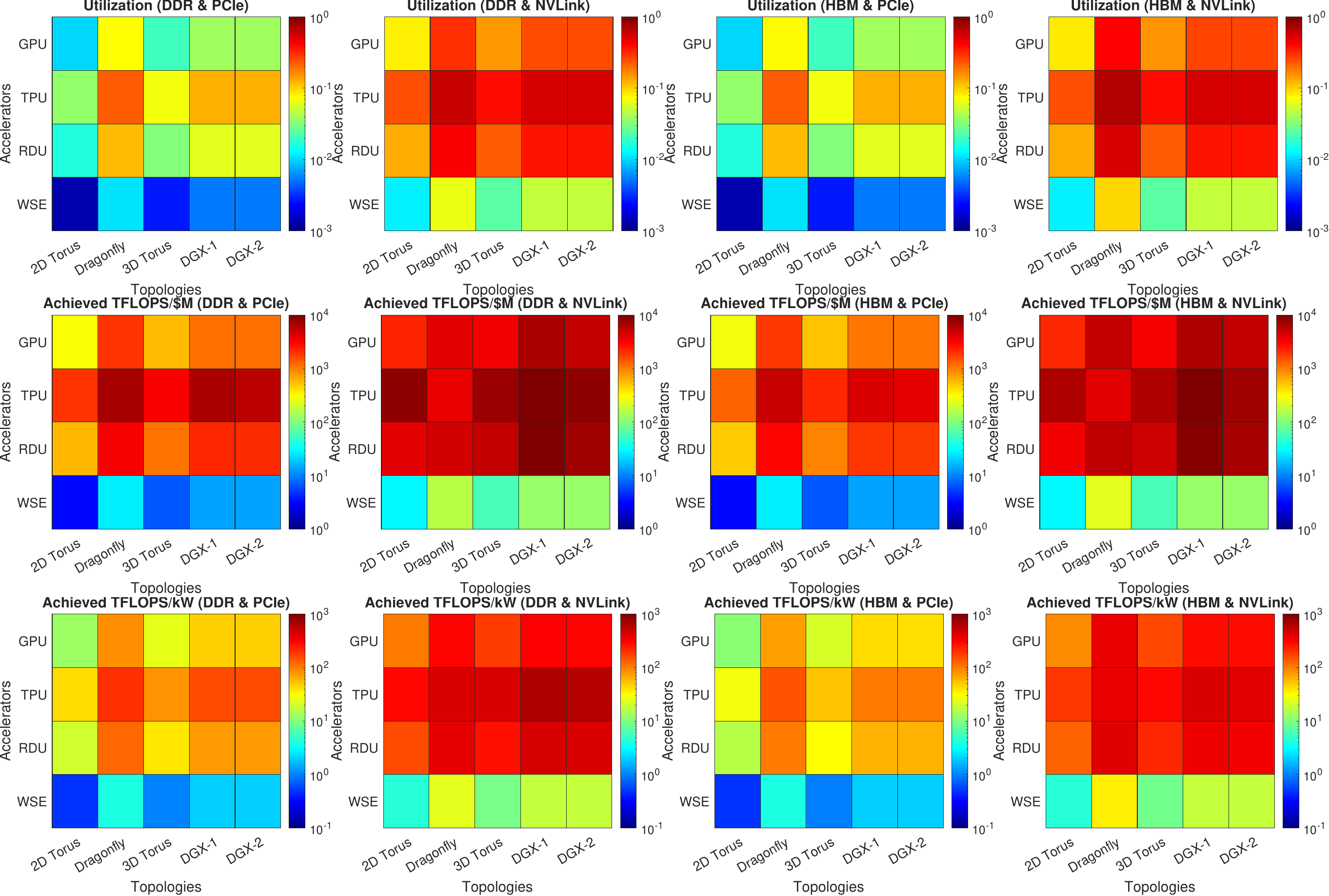}
    \vspace{-10pt}
    \caption{The heat map shows utilization, power efficiency, and cost efficiency of 793B DLRM on 1024 accelerators with different interconnection topologies and memory/network technologies.}
    \label{DLRM_heatmap}
    \vspace{-10pt}
\end{figure*}

\begin{figure*}[t!]
    \centering
    \includegraphics[width=\linewidth]{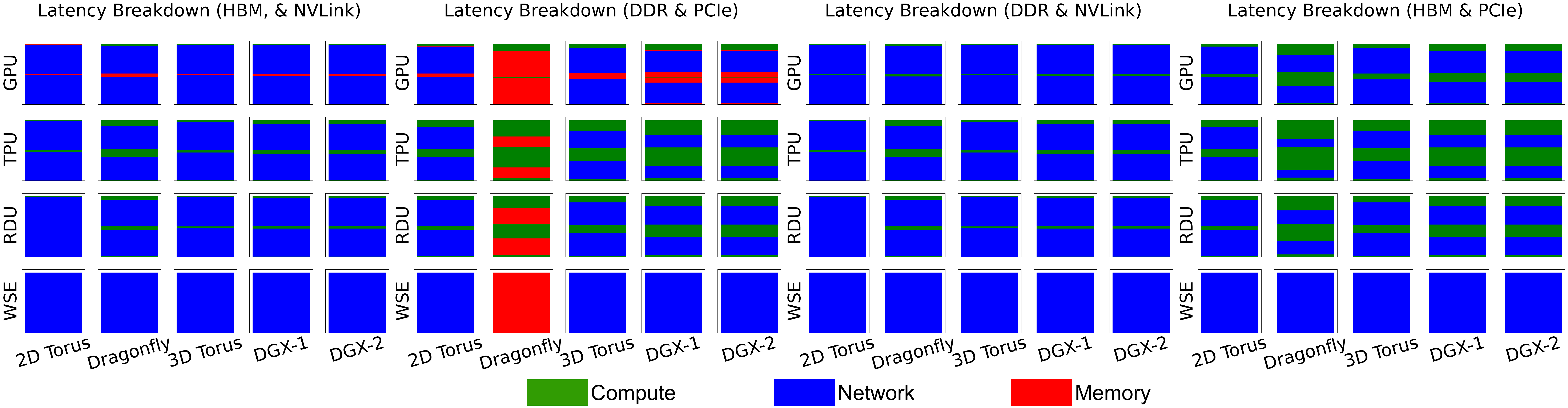}
    \vspace{-10pt}
    \caption{The plot shows the breakdown of compute/memory/network latency of 793B DLRM on 1024 accelerators with different interconnection topologies and memory/network technologies.}
    \label{profiler_DLRM}
    \vspace{-10pt}
\end{figure*}

\begin{figure*}[t!]
    \centering
    \includegraphics[width=\linewidth]{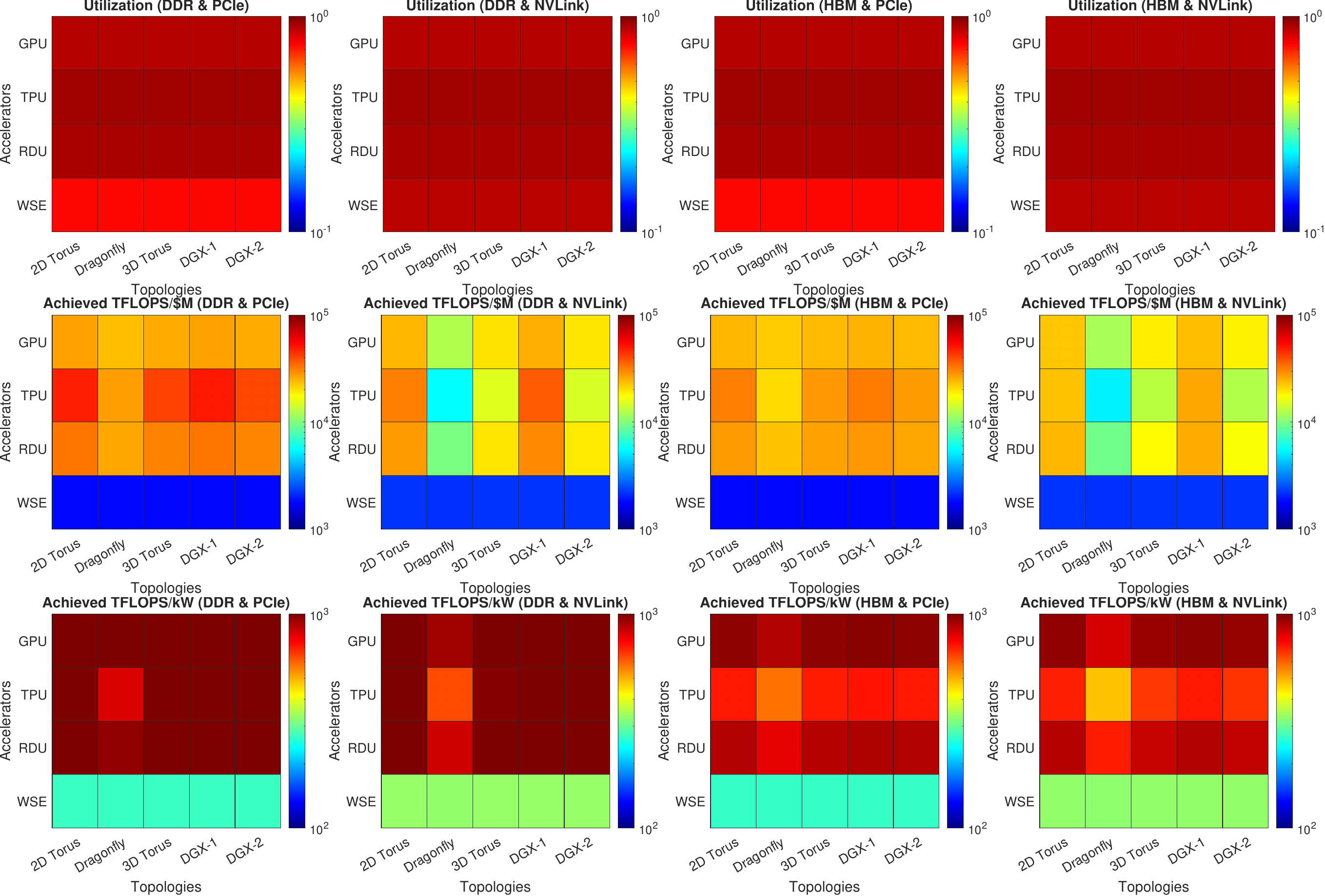}
    \vspace{-10pt}
    \caption{The heat map shows utilization, power efficiency, and cost efficiency of 5M$^2$ HPL on 1024 accelerators with different interconnection technologies and memory/network technologies.}
    \label{HPL_heatmap}
    \vspace{-10pt}
\end{figure*}

\begin{figure*}[t!]
    \centering
    \includegraphics[width=\linewidth]{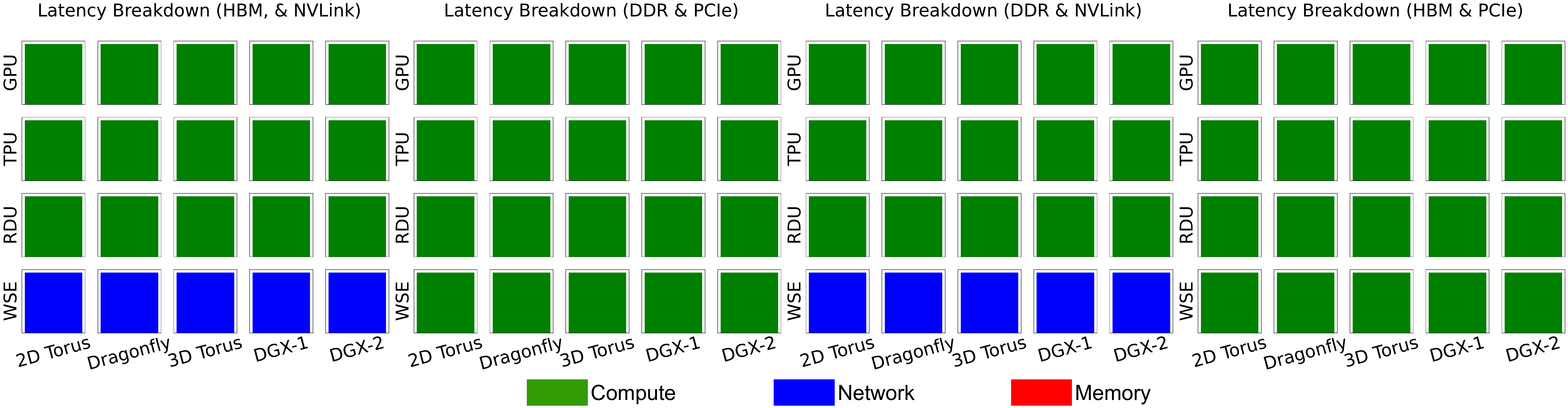}
    \vspace{-10pt}
    \caption{The plot shows the breakdown of compute/memory/network latency of 5M$^2$ on 1024 accelerators with different interconnection topologies and memory/network technologies.}
    \label{profiler_HPL}
    \vspace{-10pt}
\end{figure*}

\begin{figure*}[t!]
    \centering
    \includegraphics[width=\linewidth]{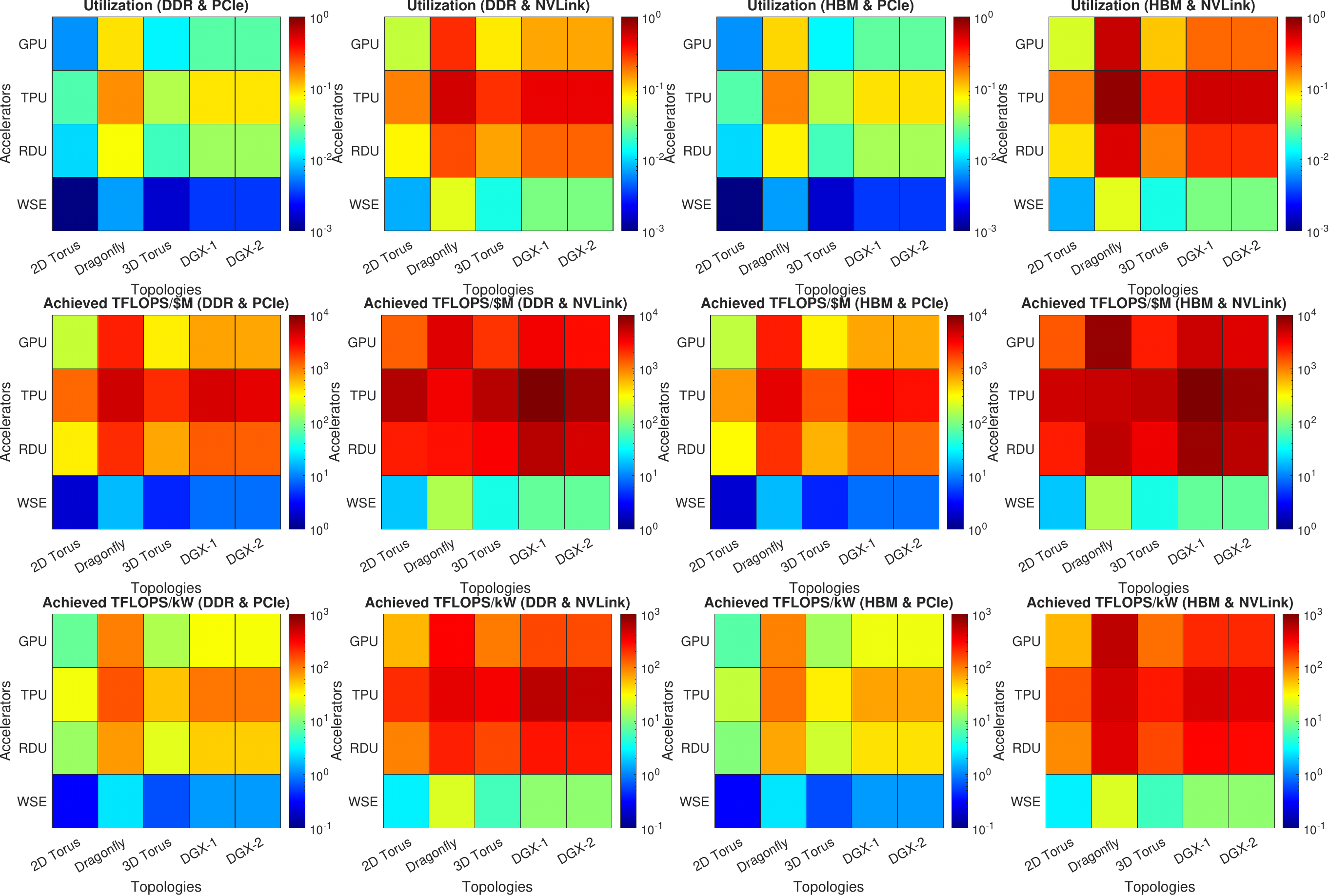}
    \vspace{-10pt}
    \caption{The heat map shows utilization, power efficiency, and cost efficiency of 1T-point FFT on 1024 accelerators with different interconnection topologies and memory/network technologies.}
    \label{FFT_heatmap}
    \vspace{-10pt}
\end{figure*}

\begin{figure*}[t!]
    \centering
    \includegraphics[width=\linewidth]{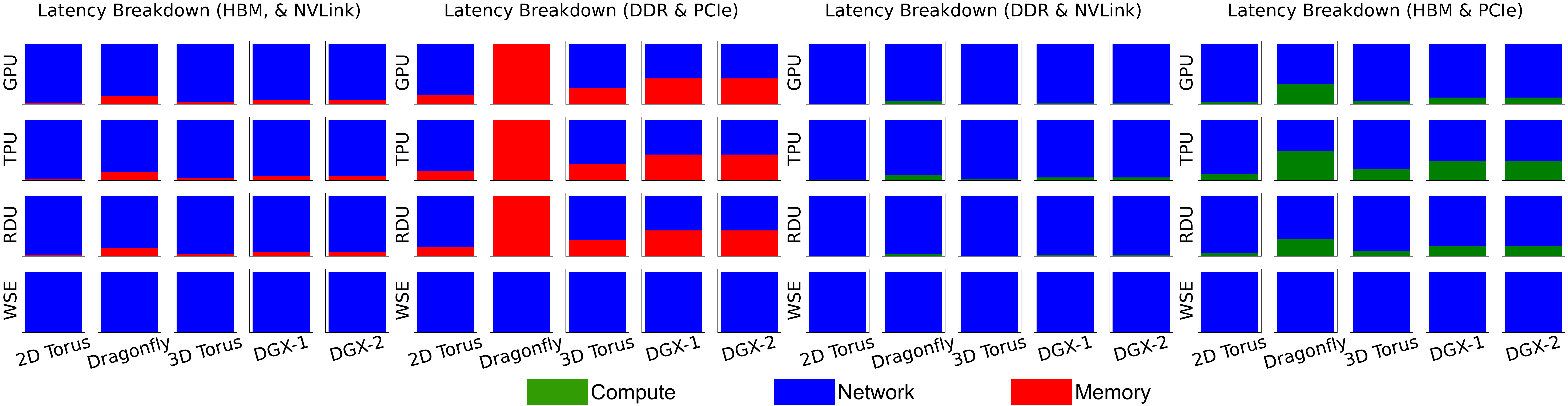}
    \vspace{-10pt}
    \caption{The plot shows the breakdown of compute/memory/network latency of 1T-point FFT on 1024 accelerators with different interconnection topologies and memory/network technologies.}
    \label{profiler_FFT}
    \vspace{-10pt}
\end{figure*}

\subsection{Design Space Exploration}
In this section, we perform a design space exploration of system parameters by exploring off-the-shelf hardware components which include various accelerator chips, memory technologies, and interconnect technologies.
For accelerators, we study the four chips in Table~\ref{chips} from different vendors.

GPUs and TPUs are non-dataflow architectures that execute in a kernel-by-kernel fashion.
RDUs are dataflow architectures that execute in the dataflow fashion.
WSEs are wafer-scale architectures that have orders of magnitude larger on-chip compute throughput and SRAM capacity than the above three architectures.
For memory technologies, we consider DDR4~\cite{ddr_pcie} with 200GB/s bandwidth and HBM3~\cite{isaev2023calculon} with 3000GB/s bandwidth.
For interconnect technologies, we consider PCIe Gen 4~\cite{ddr_pcie} with 25GB/s bandwidth and NVLink4~\cite{nvlink} with 900GB/s bandwidth.
In addition to performance, we also estimate the price and power of each system design point.
For accelerators, we collect the power of different chips from~\cite{isaev2023calculon, h100, jouppi2023tpu, prabhakar2022sambanova, wse}.
Figure~\ref{cost_function} plots the price of the accelerators versus their compute throughput.
We fit a polynomial regression curve to the data points, which shows a superlinear relationship between compute throughput and silicon power.
The price versus compute throughput also follows a similar trend.
Ultimately, these relationships imply that building large chips will incur larger price and power penalties.
In addition, we collect the price and power for memory technologies from~\cite{isaev2023calculon, jouppi2021ten} and
for interconnect technologies from~\cite{won2021exploring, nvlink_price_power}.

In our design space exploration, we target four workloads and 80 system specifications.
The four workloads are GPT3 1T~\cite{brown2020language}, 793B deep learning recommendation model (DLRM)~\cite{10.1145/3470496.3533727, gomez2015netflix}, 5M$^2$ matrix high-performance LINPACK (HPL)~\cite{hpl}, and 1T-point FFT~\cite{jung2016parallel, simonovic2018cosmological}.
The 80 system setups come from the cartesian combination of the four accelerators in Table~\ref{chips}, five different interconnection topologies: 2D torus, 3D torus, dragonfly~\cite{kim2008technology}, DGX-1~\cite{dgx_1}, and DGX-2~\cite{li2019evaluating}, and four combinations of memory and interconnect technologies: DDR \& PCIe, DDR \& NVLink, HBM \& PCIe, and HBM \& NVLink.
All system setups are configured to 1024 accelerators to form a homogeneous system.

\subsubsection{LLM}
For GPT3 1T~\cite{brown2020language}, we measured and cross-validated the throughput utilization, cost efficiency, and power efficiency for each system setup in the design space, as shown in the heat map in Figure~\ref{LLM_heatmap}.
We also measured the latency breakdown in terms of compute/memory/network for each design point, as shown in Figure~\ref{profiler_LLM}.
We make several key observations from studying the entire system design space for LLM workloads:
\begin{itemize}

\item \textit{Dataflow architectures are more performant and efficient than non-dataflow architectures}: RDUs achieve $1.52\times$ utilization compared to GPUs/TPUs, $1.59\times$ cost efficiency compared to GPUs/TPUs, and $1.6\times$ power efficiency compared to GPUs/TPUs.

\item \textit{Non-dataflow architectures require fast memory technologies to achieve high performance, while dataflow architectures do not rely on fast memory technologies to achieve high performance}: RDUs with HBM achieve the same utilization compared to RDUs with DDR, while GPUs/TPUs with HBM achieve $1.66\times$ utilization compared to GPUs/TPUs with DDR.

\item \textit{For fast interconnect technologies, simple topologies made of torus and switches such as 2D torus/3D torus/DGX-1/DGX-2 can achieve high performance, while for slow interconnect technologies, fully-connected topologies like dragonfly increase performance but come with significant cost/power overheads}: The dragonfly topology with NVLink achieves the same utilization as simple topologies with NVLink, while the dragonfly topology with PCIe achieves $1.21\times$ utilization compared to simple topologies such as 2D torus/3D torus/DGX-1/DGX-2 with PCIe, but the dragonfly topology is $0.51\times$ cost efficient and $0.91\times$ power efficient compared to simple topologies.

\item \textit{Wafer-scale accelerators do not benefit from fast memory technologies due to their high on-chip SRAM capacity but need fast interconnect technologies to keep up with on-chip compute throughput demand}: WSEs with HBM achieve the same utilization as WSEs with DDR, while WSEs with NVLink achieve $5.15\times$ utilization compared to WSEs with PCIe.

\item \textit{Wafer-scale accelerators have low cost/power efficiency due to the superlinear relationship between silicon area and cost/power}: WSEs are $0.06\times$ cost efficient and $0.2\times$ power efficient compared to GPUs/TPUs/RDUs.

\end{itemize}

\subsubsection{DLRM}
For 793B DLRM~\cite{10.1145/3470496.3533727, gomez2015netflix}, we measured and cross-validated the throughput utilization, cost efficiency, and power efficiency for each system setup in the design space, as shown in the heat map in Figure~\ref{DLRM_heatmap}.
We also measured the latency breakdown in terms of compute/memory/network for each design point, as shown in Figure~\ref{profiler_DLRM}.
We make several key observations from studying the entire system design space for LLM workloads:
\begin{itemize}

\item \textit{Fast interconnect technologies or fully-connected topologies like dragonfly are needed to achieve high performance and cost/power efficiency due to the intensive all-to-all communication in the DLRM workload}: Systems with NVLink achieve $6.3\times$ utilization, $4.31\times$ cost efficiency, and $6.04\times$ power efficiency compared to systems with PCIe, and systems with the dragonfly topology achieve $2.51\times$ utilization, $1.66\times$ cost efficiency, and $2.13\times$ power efficiency compared to systems with simple topologies.

\item \textit{Slower accelerators achieve higher utilization and cost/power efficiency when compared to faster accelerators. Slower chips waste less compute throughput in the presence of an intensive network communication}: TPUs achieve $4.43\times$ utilization, $9.45\times$ cost efficiency, and $5.24\times$ power efficiency compared to other accelerators.

\item \textit{Wafer-scale accelerators achieve significantly lower utilization and cost/power efficiency because compute throughput is wasted in the presence of an intensive network communication}: WSEs achieve $0.1\times$ utilization, $0.01\times$ cost efficiency, and $0.04\times$ power efficiency compared to other accelerators.

\end{itemize}

\subsubsection{HPL}
HPLinpack (HPL) is used to evaluate the performance of supercomputers by solving a dense linear system~\cite{hpl}.
For 5M$^2$ HPL, we measured and cross-validated the throughput utilization, cost efficiency, and power efficiency for each system setup in the design space, as shown in the heat map in Figure~\ref{HPL_heatmap}.
We also measured the latency breakdown in terms of compute/memory/network for each design point, as shown in Figure~\ref{profiler_HPL}.
We make several key observations from studying the entire system design space for LLM workloads:
\begin{itemize}

\item \textit{Fast interconnect technologies, fast memory technologies, and fully-connected topologies like dragonfly are not needed to achieve high utilization since the HPL workload is dense}: All systems setups achieve high utilization.

\item \textit{Wafer-scale accelerators achieve significantly lower cost/power efficiency due to the superlinear relationship between silicon area and cost/power}: WSEs achieve $0.09\times$ cost efficiency, and $0.33\times$ power efficiency compared to GPUs/TPUs/RDUs.

\end{itemize}

\subsubsection{FFT}
For 1T-point FFT~\cite{jung2016parallel, simonovic2018cosmological}, we measured and cross-validated the throughput utilization, cost efficiency, and power efficiency for each system setup in the design space, as shown in the heat map in Figure~\ref{FFT_heatmap}.
We also measured the latency breakdown in terms of compute/memory/network for each design point, as shown in Figure~\ref{profiler_FFT}.
We make several key observations from studying the entire system design space for LLM workloads:
\begin{itemize}

\item \textit{Fast interconnect technologies or fully-connected topologies like dragonfly are necessary to achieve high performance and cost/power efficiency due to the intensive all-to-all communication in the workload}: Systems with NVLink achieve $7.02\times$ utilization, $4.81\times$ cost efficiency, and $6.73\times$ power efficiency compared to systems with PCIe, and systems with the dragonfly topology achieve $3.22\times$ utilization, $2.13\times$ cost efficiency, and $2.74\times$ power efficiency compared to systems with simple topologies.

\item \textit{Slower accelerators achieve higher utilization and cost/power efficiency compared to faster accelerators since slower chips waste less compute throughput in the presence of an intensive network communication}: TPUs achieve $5.11\times$ utilization, $10.92\times$ cost efficiency, and $6.06\times$ power efficiency compared to other accelerators.

\item \textit{Wafer-scale accelerators have significantly lower utilization and cost/power efficiency since they waste compute throughput in the presence of an intensive network communication}: WSEs achieve $0.09\times$ utilization, $0.01\times$ cost efficiency, and $0.03\times$ power efficiency compared to other accelerators.

\end{itemize}


\section{Dataflow Mappings Exploration using DFModel}
\begin{figure}[t!]
  \centering
  \includegraphics[width=\linewidth]{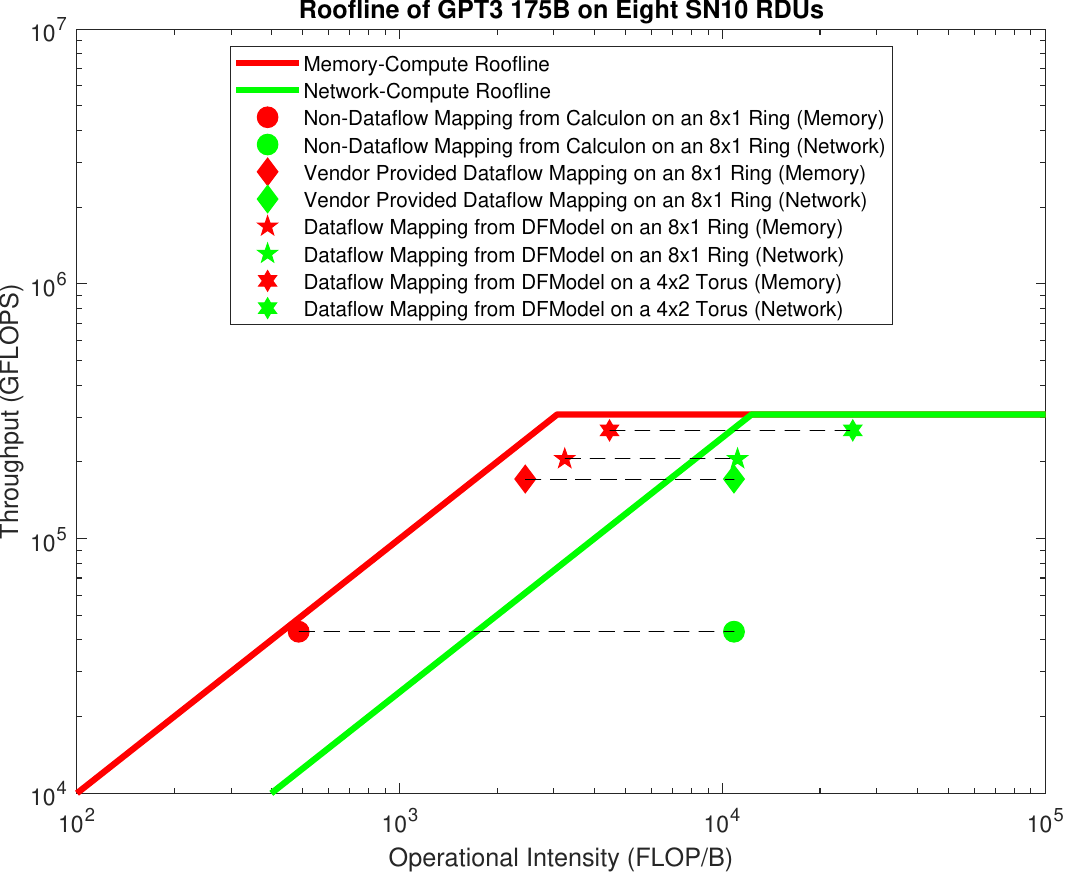}
  \vspace{-10pt}
  \caption{The hierarchical roofline plot shows the RDU chip with DDR memory and PCIe network.
  Four different mappings of GPT3 175B model achieve different throughput.
  Each mapping has two operational intensities (OIs) with respect to memory and network.
  The two OIs have the same achieved throughput.}
  \label{roofline}
  \vspace{-10pt}
\end{figure}

The previous section on design space exploration shows that DFModel can accurately capture the characteristics of different workloads and different systems, give an estimation for performance, and analyze system-level bottlenecks.
In this case study, we want to demonstrate DFModel's capability of finding an optimal dataflow mapping, which previous performance models cannot achieve.
In the experiment, we use DFModel to model GPT3 175B~\cite{brown2020language} on eight SambaNova SN10 RDUs~\cite{prabhakar2021sambanova, prabhakar2022sambanova}.
Each RDU has 307.2TFLOPS of half-precision floating-point compute throughput and 320MB of SRAM on-chip.
Each chip is equipped with large-capacity DDR memory with 200GB/s bandwidth and PCIe interconnect with 25GB/s bandwidth.
The default topology is an $8\times1$ ring with TP = 8, PP = 1, and DP = 1.
In this section, we will analyze several GPT3 mappings from least performant (non-dataflow mapping) to most performant (DFModel mapping), guided by the hierarchical roofline plot in Figure~\ref{roofline}.

\subsection{Non-Dataflow Mapping on an $8\times1$ Ring}
Previous models like Calculon~\cite{isaev2023calculon} do not model intra-chip dataflow and their mappings default to the kernel-by-kernel execution style.
When we generate the mapping using Calculon~\cite{isaev2023calculon}, it is heavily memory-bound and achieves low utilization (shown in Figure~\ref{roofline}), since memory traffic is heavy in the kernel-by-kernel execution of non-dataflow mappings.

\begin{table}[b!]
    \centering
    \vspace{-10pt}
    \caption{Comparison of four different mappings.}
    \begin{tabu}{ p{4cm}p{1.4cm}p{0.8cm}p{0.8cm} }

    \rowfont{\bfseries} \multirow{2}{*}{Mapping} & \multirow{2}{*}{Topology} & Stepwise & Accum. \\
    \rowfont{\bfseries} & & Speedup & Speedup\\
    \midrule
    
    Non-Dataflow Mapping~\cite{isaev2023calculon} & $8\times1$ Ring & $1\times$ & $1\times$ \\
    
    Vendor Provided Dataflow Mapping & $8\times1$ Ring & $4.05\times$ & $4.05\times$\\

    DFModel Dataflow Mapping & $8\times1$ Ring & $1.19\times$ & $4.8\times$\\

    DFModel Dataflow Mapping & $4\times2$ Torus & $1.28\times$ & $6.13\times$\\
    
    \end{tabu}
    \label{overall_performance}
\end{table}

\subsection{Vendor Provided Dataflow Mapping on an $8\times1$ Ring}
The vendor provided mapping from SambaNova~\cite{prabhakar2021sambanova} uses four partitions to map a layer of a GPT3 model to the system: Partition 1$\rightarrow$\{Q,K,V\}; Partition 2$\rightarrow$\{MHA1,Softmax,MHA2,Proj\}; Partition 3$\rightarrow$\{FFN0\}; Partition 4$\rightarrow$\{FFN1,Add\}.
We model the vendor provided mapping in DFModel and compare the modeled performance against the measured performance from the industrial system.
The modeled performance is within 3\% of the measured performance.
This small amount of error validates DFModel's capability of accurately modeling dataflow mappings.
Compared to the non-dataflow mapping from previous models, dataflow mappings significantly increase performance by reducing memory traffic (shown in Figure~\ref{roofline}).
Therefore, the dataflow mapping is less memory-bound and achieves $4.05\times$ speedup compared to the non-dataflow mapping.

\subsection{DFModel-Optimized Mapping on an $8\times1$ Ring}
The DFModel-optimized mapping beats the vendor provided mapping by $1.2\times$.
The difference between the DFModel-optimized mapping and the vendor provided dataflow mapping lies in Partitions 2 and 3.
Partitions 2 and 3 in the vendor provided mapping take a long execution time due to the network-intensive all-reduce from the Proj kernel in Partition 2 and the compute-intensive GEMM operation from the FFN0 kernel in Partition 3.
However, the DFModel-optimized mapping intelligently places the Proj and the FFN0 kernel into the same Partition 3 to overlap the two long latencies.
As a result, Partition 2 in the DFModel-optimized mapping becomes much faster while Partition 3 retains similar performance as before.
Overall, the DFModel-optimized mapping achieves $1.19\times$ speedup compared to the vendor provided dataflow mapping.

\subsection{DFModel-Optimized Mapping on a $4\times2$ Torus}
When analyzing the DFModel-optimized mapping from the previous subsection using the roofline analysis~\cite{williams2009roofline}, it is still network-bound (shown in Figure~\ref{roofline}).
To further improve performance, DFModel suggests a mapping on a $4\times2$ torus which is compute-bound and achieves higher throughput due to TP decreasing from 8 to 4.
As fewer chips are used for sharding, the per-chip compute FLOPs increase while the communication size in the network remains the same, increasing OI (shown in Figure~\ref{roofline}).
Therefore, the mapping changes from network-bound to compute-bound.
The DFModel-optimized mapping on a $4\times2$ torus achieves $1.28\times$ speedup compared to the DFModel-optimized mapping on an $8\times1$ ring.
Overall, Table~\ref{overall_performance} compares the four mappings discussed above.
The DFModel-optimized mapping beats non-dataflow mappings by $6.13\times$ and beats the vendor provided dataflow-mapping by $1.52\times$.

\begin{figure}[t!]
  \centering
  \includegraphics[width=\linewidth]{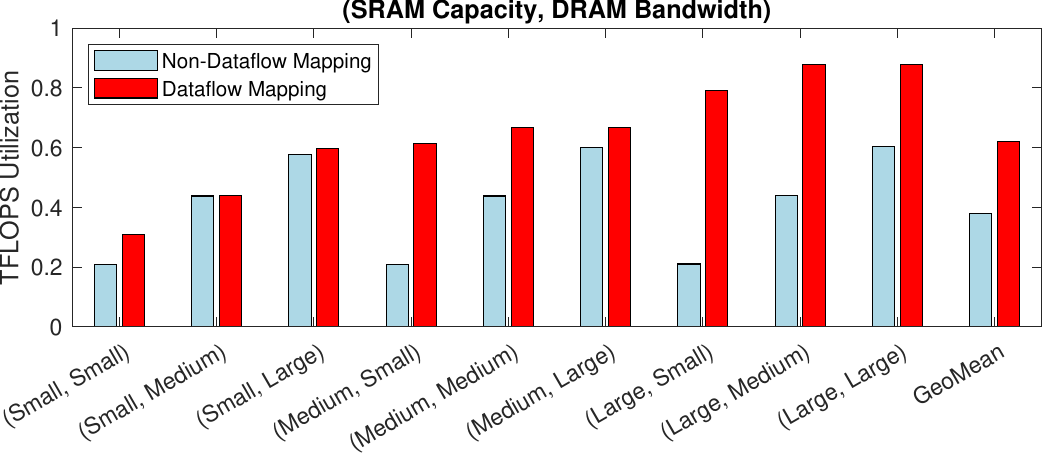}
  \vspace{-10pt}
  \caption{Comparison of dataflow mappings against non-dataflow mappings given nine different memory system setups.
  Dataflow mapping performance provides an upper bound ($1.63\times$) of non-dataflow mapping performance.
  Dataflow mappings achieve high performance when SRAM capacity is large.
  Non-dataflow mappings achieve high performance when the DRAM bandwidth is large.}
  \label{sram_dram}
  \vspace{-10pt}
\end{figure}

\subsection{Dataflow Mappings vs. Non-Dataflow Mappings}
We explore the SRAM capacity and DRAM bandwidth design space to study the relationship between dataflow mappings and the memory system.
In this experiment, on an accelerator with 300TFLOPS of compute throughput, we sweep three SRAM capacity points: 150MB (small), 300MB (medium), 500MB (large), and three DRAM bandwidth points: 100GB/s (small), 300GB/s (medium), 600GB/s (large).
We run DFModel on GPT3 175B model~\cite{brown2020language} on eight accelerators connected in a $4\times2$ torus.
Figure~\ref{sram_dram} shows the comparison of the TFLOPS utilization of dataflow mappings vs. non-dataflow mappings.
We have several key observations:
\begin{itemize}
\item Large SRAM capacity is needed for dataflow mappings to attain high performance since large SRAM capacity enables more fusion and less DRAM traffic.

\item Large DRAM bandwidth is needed for non-dataflow mappings to attain high performance since non-dataflow mappings by default incur heavy DRAM traffic.

\item Dataflow mapping performance is an upper bound of non-dataflow mapping performance ($1.63\times$).

\end{itemize}

\section{Case Studies using DFModel}
In this section, we use DFModel to explore the dataflow mapping performance of different workloads and systems.

\subsection{DFModel for LLM Serving}

LLM serving consists of a prefill phase and a decode phase~\cite{zhong2024distserve}.
In prefill, the input prompt is processed in one pass to fill the KV cache and generate one token.
It resembles the forward pass of the training process.
Users care about the time to first token (TTFT) metric which measures the latency of the entire pass of the prefill phase.
In addition, server providers care about the prefill throughput in tokens/second which measures the system-level throughput of serving different prefills for different users.
In decode, text is generated token by token autoregressively, in which each generated token is fed back into the model and combined with the KV cache to generate the next token.
Users care about the time per output token (TPOT) metric which measures the latency of the entire pass of the decode phase to generate one token.
In addition, server providers care about the decode throughput in tokens/second which measures the system-level throughput of serving different decodes for different users.

\begin{figure*}[t!]
  \centering
  \includegraphics[width=\linewidth]{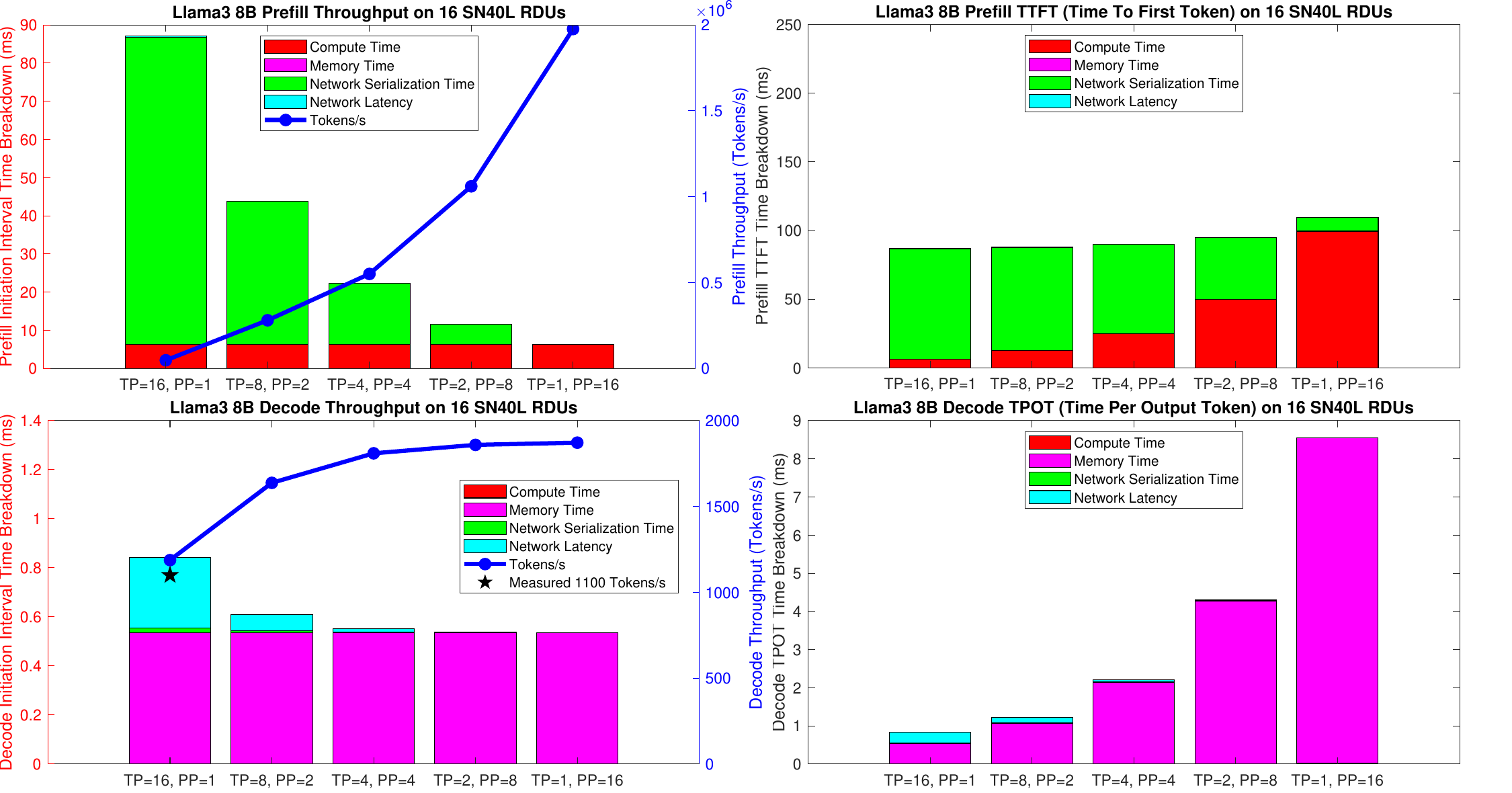}
  \vspace{-10pt}
  \caption{Serving Llama3 8B on 16 SN40L RDUs.
  TTFT (time to first token), and throughput for the prefill phase, TPOT (time per output token), and the decode phase throughput.}
  \label{prefill_decode}
  \vspace{-10pt}
\end{figure*}

In the experiment, we model the inference of Llama3 8B model~\cite{dubey2024llama3herdmodels} on 16 SambaNova SN40L RDUs~\cite{prabhakar2024sambanova}.
The system is connected in a torus topology.
Each chip contains 640TFLOPS BF16 compute throughput, 520MB of SRAM, 25GB/s network bandwidth, and 150ns network latency.
Figure~\ref{prefill_decode} shows TTFT, TPOT, and throughput for both prefill and decode phases for different combinations of TP and PP.
First, we validate the modeled decoding throughput of 1188 tokens/s for TP=16 and PP=1 against the measured 1100 tokens/s.
The error margin is only 8\%.
We then conduct the full design space exploration and we have four observations:
\begin{itemize}
    \item Increasing TP decreases latency like TTFT and TPOT, but the system-level throughput is decreased.
    \item Increasing PP increases the system-level throughput, but the latency like TTFT and TPOT is decreased.
    \item In the prefill phase, most time is spent on network serialization and compute.
    \item In the decode phase, most time is spent on memory and network latency.
\end{itemize}

\subsection{DFModel for Speculative Decoding}
Speculative decoding~\cite{leviathan2023fast} is a technique to accelerate a large, slow language model (target model) serving by using a small, fast model (draft model). The small model generates a sequence of tokens verified by the large model.
A key parameter in speculative decoding is window size (K), representing the number of tokens generated by the draft model in each step.
Larger window sizes improve speedup potential but require high acceptance rates, which measure the proportion of proposed tokens that the verification model agrees with.
SpecInfer~\cite{10.1145/3620666.3651335} advances speculative decoding by generating a tree of tokens with a total of 2$^K$ tokens compared to the conventional K tokens.
This approach significantly improves the acceptance rate due to the high path diversity. However, imposes speed challenges for the draft model since it requires it to generate an exponential amount of tokens.

\begin{figure*}[t!]
  \centering
  \includegraphics[width=\linewidth]{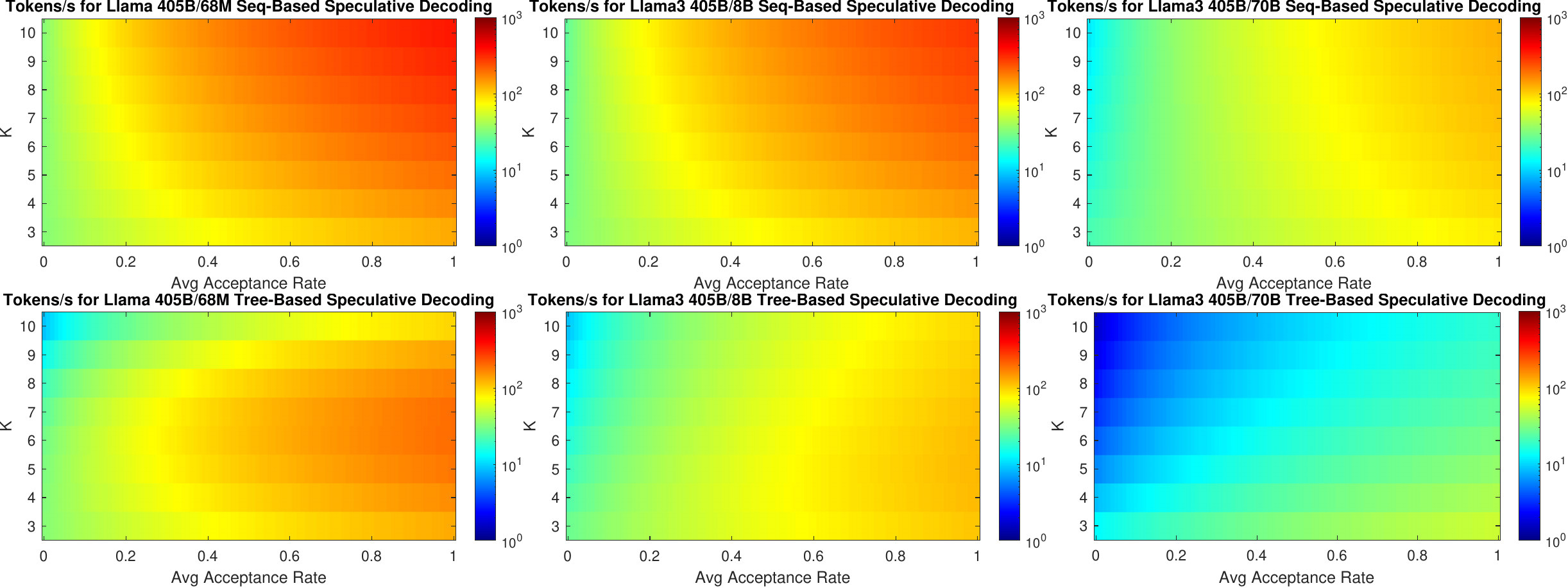}
  \vspace{-10pt}
  \caption{Two speculative decoding schemes are shown: sequence-based and tree-based.
  We sweep the acceptance rate and the window size and plot the corresponding achieved tokens per second.}
  \label{spec_decoding}
  \vspace{-10pt}
\end{figure*}

In this experiment, we use 16 SambaNova SN40L RDUs~\cite{prabhakar2024sambanova} as the serving system.
The study shows two speculative decoding schemes (sequence-based and tree-based) in which one of Llama 68M, Llama3 8B, or Llama3 70B is used as the draft model, and Llama3 405B is used as the target model.
First, we validate the accuracy of DFModel by comparing the decoding throughput of the 70B (paired with 8B draft model) and 405B (paired with 8B draft model) with the measured throughput from~\cite{prabhakar2024sambanova}.
The result is shown in Figure~\ref{overall_comparison} and the error margin is 12\%.
We sweep the acceptance rate and the window size and report the achieved tokens per second for decoding, as shown in Figure~\ref{spec_decoding}.
We have several observations:
\begin{itemize}
    \item Tree-based speculative decoding needs the smallest 68M model as the draft model and a short window size since the overheads are too high when generating an exponential amount of tokens.
    \item For sequence-based speculative decoding, small 68M and medium 8B draft models are desirable while the large 70B draft model has too much overheads.
    \item For sequence-based speculative decoding, increasing both window size and acceptance rate improves performance.
\end{itemize}

\subsection{DFModel for 3D Memory}
The 3D-stacked system has been popular recently to break the challenging memory wall problem for AI workloads~\cite{gao2017tetris, rezaei2023smart, lee2021task, hadidi2017demystifying, kim2016neurocube}.
The current 2.5D system integrates an HBM with a logic die on the same silicon interposer, and the memory bandwidth is proportional to the die perimeter.
In the future 3D system, an HBM is stacked on top of a logic die, and the memory bandwidth is proportional to the die area, which gives an order of magnitude more memory bandwidth when compared to the 2.5D system.

\begin{figure}[t!]
  \centering
  \includegraphics[width=\linewidth]{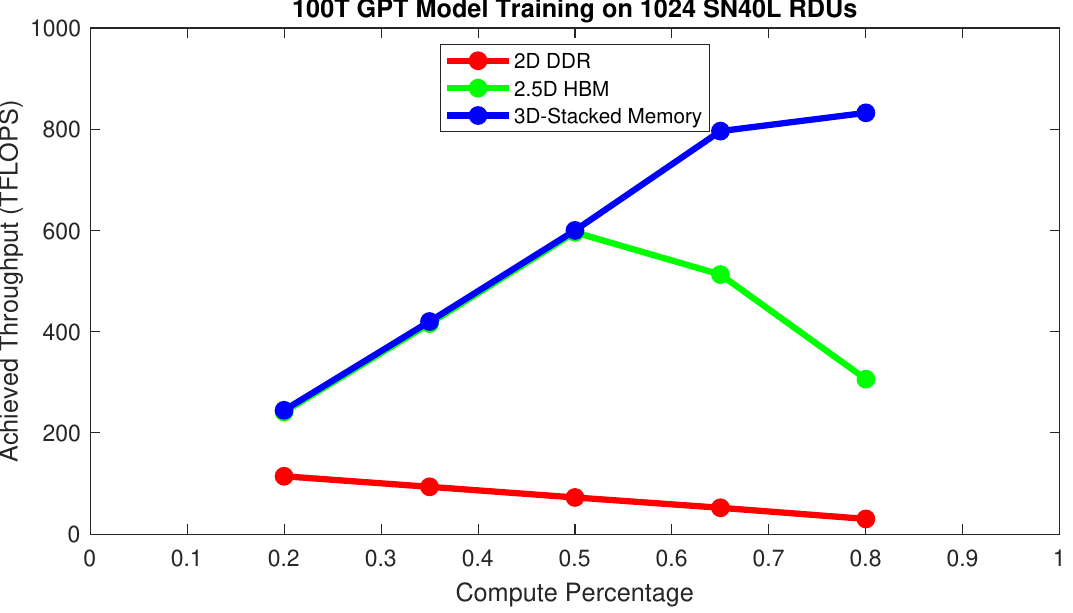}
  \vspace{-10pt}
  \caption{The achieved throughput of different compute percentages when given different off-chip memory technologies.
  As off-chip memory bandwidth increases, the higher percentage of the chip area can be computed and the lower percentage of the chip area can be SRAM memory to achieve higher throughput. 
  }
  \label{memory_3d}
  \vspace{-10pt}
\end{figure}

In this experiment, we use 1024 SambaNova SN40L RDUs with three different off-chip memory technologies: 2D DDR w/ 100GB/s, 2.5D HBM w/ 1TB/s, and 3D-stacked memory w/ 100TB/s~\cite{dally2022nvidia}.
We use a projected 100T future GPT model following the scaling law from~\cite{narayanan2021efficient}.
Each RDU chip is packed with 2080 iso-area units with 1040 compute units and 1040 memory units.
We fix the total number of units to 2080 and sweep the compute tile percentage from 20\% to 80\%.
Figure~\ref{memory_3d} shows the achieved throughput for training the 100T GPT model under different scenarios.
We have several observations:
\begin{itemize}
    \item For the low-bandwidth 2D DDR memory, more on-chip memory is needed to avoid being memory-bound and achieve high performance.
    \item For the medium-bandwidth 2.5D HBM memory, a balanced design of an equal amount of compute and memory on-chip is preferred.
    \item For the high-bandwidth 3D-stacked memory, more compute is preferred to give a chip a higher compute throughput upper-bound. The chip can tolerate low on-chip memory density due to the ultra-fast 3D memory.
\end{itemize}

\section{Related Work}
\paragraph{Single chip mapping}
Scheduling workloads to a single dataflow chip has been a popular research area.
Previous works like~\cite{zhao2020towards} propose a modulo scheduling technique to schedule a hyperblock of code to a coarse-grain reconfigurable array (CGRA).
Works like~\cite{nowatzki2013general, nowatzki2018hybrid} use integer-linear programming techniques to map a given workload to a CGRA.
MapZero~\cite{kong2023mapzero} designs a compiler inspired by reinforcement learning and Monte-Carlo tree search to map a dataflow graph to a CGRA.
HierCGRA~\cite{chenhiercgra} instead replies on graph homomorphism algorithms to generate mappings for a CGRA.
APEX~\cite{melchert2023apex} uses frequent subgraph analysis to automatically generate the processing element (PE) of a CGRA for a specific application domain.
SARA~\cite{zhang2021sara} proposes a novel mapping scheme based on heuristics and linear programming to target single-chip large-scale reconfigurable dataflow architecture (RDA).
LLM-Viewer~\cite{yuan2024llm} analyzes the LLM inference performance of a GPU.
Timeloop~\cite{parashar2019timeloop}, CoSA~\cite{huang2021cosa}, and Explainable-DSE~\cite{dave2023explainable} optimize single kernels with nested loops on a single accelerator.
FAST~\cite{zhang2022full} and Mind the Gap~\cite{mindthegap} explore kernel fusion on a single accelerator.
Compared to DFModel, the above frameworks focus on mapping workloads to a single chip, and inter-chip dataflow mappings are not covered by these frameworks.

\paragraph{Distributed system mapping}
DFModel models general dataflow which can deal with arbitrary dataflow graphs and system specifications whereas previous distributed mapping works can only process domain-specific workloads.
FlexFlow~\cite{jia2019beyond}, PipeMare~\cite{yang2021pipemare}, Alpa~\cite{zheng2022alpa}, and Megatron-LM~\cite{narayanan2021efficient} optimize inter-chip dataflow mappings across accelerators, but do not consider intra-chip dataflow mappings and system design space exploration.
Calculon~\cite{isaev2023calculon}, Rail-Only~\cite{wang2023optimized}, ASTRA-sim~\cite{rashidi2020astra}, and LLMCompass~\cite{llmcompass} focus on optimizing both inter-chip dataflow mappings and system exploration, but are unable to model intra-chip dataflow mappings.
On the contrary, DFModel optimizes dataflow mappings at several hierarchies of the system including both inter-chip level and intra-chip level.
In addition, we use DFModel to explore and optimize the entire system design space.
\section{Conclusion}
This paper introduces DFModel, a modeling
framework for mapping workload dataflow graphs onto large-scale systems.
To the best of our knowledge, DFModel is the first tool to optimize dataflow mappings at several hierarchies of a distributed system including the inter-chip level and intra-chip level. DFModel does this optimization by formulating the mapping space into an optimization problem which is automatically solved by Gurobi~\cite{gurobi2022gurobi}.
DFModel is able to swiftly produce a mapping with provably optimal performance for a trillion-parameter-scale LLM onto a thousand-accelerator datacenter, exploring a design space of size $O(10^{295})$ within 20 minutes on a server with 64 CPUs.
DFModel's performance estimation is validated against previous performance models~\cite{isaev2023calculon, wang2023optimized} as well as the measured performance from industrial systems.
DFModel can be used to explore the system and algorithmic design space by considering different parallelization strategies, accelerator architectures, interconnect/memory technologies, and interconnection network topologies.
We believe DFModel will drive future research into designing large-scale systems for future workloads.

\newpage


\bibliographystyle{IEEEtranS}
\bibliography{refs}

\begin{thebibliography}{10}
\providecommand{\url}[1]{#1}
\csname url@samestyle\endcsname
\providecommand{\newblock}{\relax}
\providecommand{\bibinfo}[2]{#2}
\providecommand{\BIBentrySTDinterwordspacing}{\spaceskip=0pt\relax}
\providecommand{\BIBentryALTinterwordstretchfactor}{4}
\providecommand{\BIBentryALTinterwordspacing}{\spaceskip=\fontdimen2\font plus
\BIBentryALTinterwordstretchfactor\fontdimen3\font minus \fontdimen4\font\relax}
\providecommand{\BIBforeignlanguage}[2]{{%
\expandafter\ifx\csname l@#1\endcsname\relax
\typeout{** WARNING: IEEEtranS.bst: No hyphenation pattern has been}%
\typeout{** loaded for the language `#1'. Using the pattern for}%
\typeout{** the default language instead.}%
\else
\language=\csname l@#1\endcsname
\fi
#2}}
\providecommand{\BIBdecl}{\relax}
\BIBdecl

\bibitem{ddr_pcie}
\BIBentryALTinterwordspacing
``Performance characteristics of common transports and buses,'' 2013. [Online]. Available: \url{https://www.microway.com/knowledge-center-articles/performance-characteristics-of-common-transports-buses/}
\BIBentrySTDinterwordspacing

\bibitem{dgx_1}
\BIBentryALTinterwordspacing
``Nvidia dgx-1 with tesla v100 system architecture,'' 2017. [Online]. Available: \url{https://images.nvidia.com/content/pdf/dgx1-v100-system-architecture-whitepaper.pdf}
\BIBentrySTDinterwordspacing

\bibitem{Circe}
\BIBentryALTinterwordspacing
``Circe,'' 2018. [Online]. Available: \url{https://www.top500.org/system/179564/}
\BIBentrySTDinterwordspacing

\bibitem{hpl}
\BIBentryALTinterwordspacing
``Hpl - a portable implementation of the high-performance linpack benchmark for distributed-memory computers,'' 2018. [Online]. Available: \url{https://www.netlib.org/benchmark/hpl/}
\BIBentrySTDinterwordspacing

\bibitem{Selene}
\BIBentryALTinterwordspacing
``Selene,'' 2020. [Online]. Available: \url{https://www.top500.org/system/179842/}
\BIBentrySTDinterwordspacing

\bibitem{wse}
\BIBentryALTinterwordspacing
``Cerebras unveils wafer scale engine two (wse2): 2.6 trillion transistors, 100
\BIBentrySTDinterwordspacing

\bibitem{Tethys}
\BIBentryALTinterwordspacing
``Tethys,'' 2021. [Online]. Available: \url{https://www.top500.org/system/180034/}
\BIBentrySTDinterwordspacing

\bibitem{cufftmp}
\BIBentryALTinterwordspacing
``Multinode multi-gpu: Using nvidia cufftmp ffts at scale,'' 2022. [Online]. Available: \url{https://developer.nvidia.com/blog/multinode-multi-gpu-using-nvidia-cufftmp-ffts-at-scale/}
\BIBentrySTDinterwordspacing

\bibitem{gurobi2022gurobi}
\BIBentryALTinterwordspacing
``Gurobi optimizer reference manual, version 10.0,'' 2023. [Online]. Available: \url{https://www.gurobi.com/wp-content/plugins/hd_documentations/documentation/10.0/refman.pdf}
\BIBentrySTDinterwordspacing

\bibitem{h100}
\BIBentryALTinterwordspacing
``Nvidia h100 pcie 96 gb,'' 2023. [Online]. Available: \url{https://www.techpowerup.com/gpu-specs/h100-pcie-96-gb.c4164}
\BIBentrySTDinterwordspacing

\bibitem{nvlink_price_power}
\BIBentryALTinterwordspacing
``What is nvlink?'' 2023. [Online]. Available: \url{https://blogs.nvidia.com/blog/what-is-nvidia-nvlink/}
\BIBentrySTDinterwordspacing

\bibitem{nvlink}
\BIBentryALTinterwordspacing
``Nvlink and nvlink switch,'' 2024. [Online]. Available: \url{https://www.nvidia.com/en-us/data-center/nvlink/}
\BIBentrySTDinterwordspacing

\bibitem{ayala2022analysis}
A.~Ayala, S.~Tomov, P.~Luszczek, S.~Cayrols, G.~Ragghianti, and J.~Dongarra, ``Analysis of the communication and computation cost of fft libraries towards exascale,'' Technical Report ICL-UT-22-07. https://icl. utk. edu/files/publications/2022~…, Tech. Rep., 2022.

\bibitem{bambhaniya2024demystifying}
A.~Bambhaniya, R.~Raj, G.~Jeong, S.~Kundu, S.~Srinivasan, M.~Elavazhagan, M.~Kumar, and T.~Krishna, ``Demystifying platform requirements for diverse llm inference use cases,'' \emph{arXiv preprint arXiv:2406.01698}, 2024.

\bibitem{bang2024vtrain}
J.~Bang, Y.~Choi, M.~Kim, Y.~Kim, and M.~Rhu, ``vtrain: A simulation framework for evaluating cost-effective and compute-optimal large language model training,'' in \emph{2024 57th IEEE/ACM International Symposium on Microarchitecture (MICRO)}.\hskip 1em plus 0.5em minus 0.4em\relax IEEE, 2024, pp. 153--167.

\bibitem{brown2020language}
\BIBentryALTinterwordspacing
T.~Brown, B.~Mann, N.~Ryder, M.~Subbiah, J.~D. Kaplan, P.~Dhariwal, A.~Neelakantan, P.~Shyam, G.~Sastry, A.~Askell, S.~Agarwal, A.~Herbert-Voss, G.~Krueger, T.~Henighan, R.~Child, A.~Ramesh, D.~Ziegler, J.~Wu, C.~Winter, C.~Hesse, M.~Chen, E.~Sigler, M.~Litwin, S.~Gray, B.~Chess, J.~Clark, C.~Berner, S.~McCandlish, A.~Radford, I.~Sutskever, and D.~Amodei, ``Language models are few-shot learners,'' in \emph{Advances in Neural Information Processing Systems}, H.~Larochelle, M.~Ranzato, R.~Hadsell, M.~Balcan, and H.~Lin, Eds., vol.~33.\hskip 1em plus 0.5em minus 0.4em\relax Curran Associates, Inc., 2020, pp. 1877--1901. [Online]. Available: \url{https://proceedings.neurips.cc/paper_files/paper/2020/file/1457c0d6bfcb4967418bfb8ac142f64a-Paper.pdf}
\BIBentrySTDinterwordspacing

\bibitem{10.1145/3643479.3662055}
\BIBentryALTinterwordspacing
T.~Bui, O.~Tran, P.~Nguyen, B.~Ho, L.~Nguyen, T.~Bui, and T.~Quan, ``Cross-data knowledge graph construction for llm-enabled educational question-answering system: A case study at hcmut,'' in \emph{Proceedings of the 1st ACM Workshop on AI-Powered Q\&A Systems for Multimedia}, ser. AIQAM '24.\hskip 1em plus 0.5em minus 0.4em\relax New York, NY, USA: Association for Computing Machinery, 2024, p. 36–43. [Online]. Available: \url{https://doi.org/10.1145/3643479.3662055}
\BIBentrySTDinterwordspacing

\bibitem{chenhiercgra}
\BIBentryALTinterwordspacing
S.~Chen, C.~Cai, S.~Zheng, J.~Li, G.~Zhu, J.~Li, Y.~Yan, Y.~Dai, W.~Yin, and L.~Wang, ``Hiercgra: A novel framework for large-scale cgra with hierarchical modeling and automated design space exploration,'' \emph{ACM Trans. Reconfigurable Technol. Syst.}, vol.~17, no.~2, may 2024. [Online]. Available: \url{https://doi.org/10.1145/3656176}
\BIBentrySTDinterwordspacing

\bibitem{cho2019blueconnect}
M.~Cho, U.~Finkler, M.~Serrano, D.~Kung, and H.~Hunter, ``Blueconnect: Decomposing all-reduce for deep learning on heterogeneous network hierarchy,'' \emph{IBM Journal of Research and Development}, vol.~63, no.~6, pp. 1:1--1:11, 2019.

\bibitem{choquette2022nvidia}
J.~Choquette, ``Nvidia hopper gpu: Scaling performance,'' in \emph{2022 IEEE Hot Chips 34 Symposium (HCS)}, 2022, pp. 1--46.

\bibitem{choquette2020nvidia}
J.~Choquette and W.~Gandhi, ``Nvidia a100 gpu: Performance \& innovation for gpu computing,'' in \emph{2020 IEEE Hot Chips 32 Symposium (HCS)}, 2020, pp. 1--43.

\bibitem{dally2022nvidia}
B.~Dally, ``Insights from nvidia research,'' \url{https://www.youtube.com/watch?v=W6k2Q45nlA4}, 2022, lecture available on YouTube.

\bibitem{dao2023flashattention2fasterattentionbetter}
\BIBentryALTinterwordspacing
T.~Dao, ``Flashattention-2: Faster attention with better parallelism and work partitioning,'' 2023. [Online]. Available: \url{https://arxiv.org/abs/2307.08691}
\BIBentrySTDinterwordspacing

\bibitem{10.5555/3600270.3601459}
T.~Dao, D.~Y. Fu, S.~Ermon, A.~Rudra, and C.~R\'{e}, ``Flashattention: fast and memory-efficient exact attention with io-awareness,'' in \emph{Proceedings of the 36th International Conference on Neural Information Processing Systems}, ser. NIPS '22.\hskip 1em plus 0.5em minus 0.4em\relax Red Hook, NY, USA: Curran Associates Inc., 2024.

\bibitem{dave2023explainable}
\BIBentryALTinterwordspacing
S.~Dave, T.~Nowatzki, and A.~Shrivastava, ``Explainable-dse: An agile and explainable exploration of efficient hw/sw codesigns of deep learning accelerators using bottleneck analysis,'' in \emph{Proceedings of the 28th ACM International Conference on Architectural Support for Programming Languages and Operating Systems, Volume 4}, ser. ASPLOS '23.\hskip 1em plus 0.5em minus 0.4em\relax New York, NY, USA: Association for Computing Machinery, 2024, p. 87–107. [Online]. Available: \url{https://doi.org/10.1145/3623278.3624772}
\BIBentrySTDinterwordspacing

\bibitem{dean2012large}
J.~Dean, G.~S. Corrado, R.~Monga, K.~Chen, M.~Devin, Q.~V. Le, M.~Z. Mao, M.~Ranzato, A.~Senior, P.~Tucker, K.~Yang, and A.~Y. Ng, ``Large scale distributed deep networks,'' in \emph{Proceedings of the 25th International Conference on Neural Information Processing Systems - Volume 1}, ser. NIPS'12.\hskip 1em plus 0.5em minus 0.4em\relax Red Hook, NY, USA: Curran Associates Inc., 2012, p. 1223–1231.

\bibitem{Domahidi2013ecos}
A.~Domahidi, E.~Chu, and S.~Boyd, ``{ECOS}: {A}n {SOCP} solver for embedded systems,'' in \emph{European Control Conference (ECC)}, 2013, pp. 3071--3076.

\bibitem{dubey2024llama3herdmodels}
\BIBentryALTinterwordspacing
A.~Dubey, A.~Jauhri, A.~Pandey, A.~Kadian, A.~Al-Dahle, A.~Letman, A.~Mathur, A.~Schelten, A.~Yang, A.~Fan, A.~Goyal, A.~Hartshorn, A.~Yang, A.~Mitra, A.~Sravankumar, A.~Korenev, A.~Hinsvark, A.~Rao, A.~Zhang, A.~Rodriguez, A.~Gregerson, A.~Spataru, B.~Roziere, B.~Biron, B.~Tang, B.~Chern, C.~Caucheteux, C.~Nayak, C.~Bi, C.~Marra, C.~McConnell, C.~Keller, C.~Touret, C.~Wu, C.~Wong, C.~C. Ferrer, C.~Nikolaidis, D.~Allonsius, D.~Song, D.~Pintz, D.~Livshits, D.~Esiobu, D.~Choudhary, D.~Mahajan, D.~Garcia-Olano, D.~Perino, D.~Hupkes, E.~Lakomkin, E.~AlBadawy, E.~Lobanova, E.~Dinan, E.~M. Smith, F.~Radenovic, F.~Zhang, G.~Synnaeve, G.~Lee, G.~L. Anderson, G.~Nail, G.~Mialon, G.~Pang, G.~Cucurell, H.~Nguyen, H.~Korevaar, H.~Xu, H.~Touvron, I.~Zarov, I.~A. Ibarra, I.~Kloumann, I.~Misra, I.~Evtimov, J.~Copet, J.~Lee, J.~Geffert, J.~Vranes, J.~Park, J.~Mahadeokar, J.~Shah, J.~van~der Linde, J.~Billock, J.~Hong, J.~Lee, J.~Fu, J.~Chi, J.~Huang, J.~Liu, J.~Wang, J.~Yu, J.~Bitton, J.~Spisak, J.~Park, J.~Rocca,
  J.~Johnstun, J.~Saxe, J.~Jia, K.~V. Alwala, K.~Upasani, K.~Plawiak, K.~Li, K.~Heafield, K.~Stone, K.~El-Arini, K.~Iyer, K.~Malik, K.~Chiu, K.~Bhalla, L.~Rantala-Yeary, L.~van~der Maaten, L.~Chen, L.~Tan, L.~Jenkins, L.~Martin, L.~Madaan, L.~Malo, L.~Blecher, L.~Landzaat, L.~de~Oliveira, M.~Muzzi, M.~Pasupuleti, M.~Singh, M.~Paluri, M.~Kardas, M.~Oldham, M.~Rita, M.~Pavlova, M.~Kambadur, M.~Lewis, M.~Si, M.~K. Singh, M.~Hassan, N.~Goyal, N.~Torabi, N.~Bashlykov, N.~Bogoychev, N.~Chatterji, O.~Duchenne, O.~Çelebi, P.~Alrassy, P.~Zhang, P.~Li, P.~Vasic, P.~Weng, P.~Bhargava, P.~Dubal, P.~Krishnan, P.~S. Koura, P.~Xu, Q.~He, Q.~Dong, R.~Srinivasan, R.~Ganapathy, R.~Calderer, R.~S. Cabral, R.~Stojnic, R.~Raileanu, R.~Girdhar, R.~Patel, R.~Sauvestre, R.~Polidoro, R.~Sumbaly, R.~Taylor, R.~Silva, R.~Hou, R.~Wang, S.~Hosseini, S.~Chennabasappa, S.~Singh, S.~Bell, S.~S. Kim, S.~Edunov, S.~Nie, S.~Narang, S.~Raparthy, S.~Shen, S.~Wan, S.~Bhosale, S.~Zhang, S.~Vandenhende, S.~Batra, S.~Whitman, S.~Sootla, S.~Collot,
  S.~Gururangan, S.~Borodinsky, T.~Herman, T.~Fowler, T.~Sheasha, T.~Georgiou, T.~Scialom, T.~Speckbacher, T.~Mihaylov, T.~Xiao, U.~Karn, V.~Goswami, V.~Gupta, V.~Ramanathan, V.~Kerkez, V.~Gonguet, V.~Do, V.~Vogeti, V.~Petrovic, W.~Chu, W.~Xiong, W.~Fu, W.~Meers, X.~Martinet, X.~Wang, X.~E. Tan, X.~Xie, X.~Jia, X.~Wang, Y.~Goldschlag, Y.~Gaur, Y.~Babaei, Y.~Wen, Y.~Song, Y.~Zhang, Y.~Li, Y.~Mao, Z.~D. Coudert, Z.~Yan, Z.~Chen, Z.~Papakipos, A.~Singh, A.~Grattafiori, A.~Jain, A.~Kelsey, A.~Shajnfeld, A.~Gangidi, A.~Victoria, A.~Goldstand, A.~Menon, A.~Sharma, A.~Boesenberg, A.~Vaughan, A.~Baevski, A.~Feinstein, A.~Kallet, A.~Sangani, A.~Yunus, A.~Lupu, A.~Alvarado, A.~Caples, A.~Gu, A.~Ho, A.~Poulton, A.~Ryan, A.~Ramchandani, A.~Franco, A.~Saraf, A.~Chowdhury, A.~Gabriel, A.~Bharambe, A.~Eisenman, A.~Yazdan, B.~James, B.~Maurer, B.~Leonhardi, B.~Huang, B.~Loyd, B.~D. Paola, B.~Paranjape, B.~Liu, B.~Wu, B.~Ni, B.~Hancock, B.~Wasti, B.~Spence, B.~Stojkovic, B.~Gamido, B.~Montalvo, C.~Parker, C.~Burton, C.~Mejia,
  C.~Wang, C.~Kim, C.~Zhou, C.~Hu, C.-H. Chu, C.~Cai, C.~Tindal, C.~Feichtenhofer, D.~Civin, D.~Beaty, D.~Kreymer, D.~Li, D.~Wyatt, D.~Adkins, D.~Xu, D.~Testuggine, D.~David, D.~Parikh, D.~Liskovich, D.~Foss, D.~Wang, D.~Le, D.~Holland, E.~Dowling, E.~Jamil, E.~Montgomery, E.~Presani, E.~Hahn, E.~Wood, E.~Brinkman, E.~Arcaute, E.~Dunbar, E.~Smothers, F.~Sun, F.~Kreuk, F.~Tian, F.~Ozgenel, F.~Caggioni, F.~Guzmán, F.~Kanayet, F.~Seide, G.~M. Florez, G.~Schwarz, G.~Badeer, G.~Swee, G.~Halpern, G.~Thattai, G.~Herman, G.~Sizov, Guangyi, Zhang, G.~Lakshminarayanan, H.~Shojanazeri, H.~Zou, H.~Wang, H.~Zha, H.~Habeeb, H.~Rudolph, H.~Suk, H.~Aspegren, H.~Goldman, I.~Damlaj, I.~Molybog, I.~Tufanov, I.-E. Veliche, I.~Gat, J.~Weissman, J.~Geboski, J.~Kohli, J.~Asher, J.-B. Gaya, J.~Marcus, J.~Tang, J.~Chan, J.~Zhen, J.~Reizenstein, J.~Teboul, J.~Zhong, J.~Jin, J.~Yang, J.~Cummings, J.~Carvill, J.~Shepard, J.~McPhie, J.~Torres, J.~Ginsburg, J.~Wang, K.~Wu, K.~H. U, K.~Saxena, K.~Prasad, K.~Khandelwal, K.~Zand,
  K.~Matosich, K.~Veeraraghavan, K.~Michelena, K.~Li, K.~Huang, K.~Chawla, K.~Lakhotia, K.~Huang, L.~Chen, L.~Garg, L.~A, L.~Silva, L.~Bell, L.~Zhang, L.~Guo, L.~Yu, L.~Moshkovich, L.~Wehrstedt, M.~Khabsa, M.~Avalani, M.~Bhatt, M.~Tsimpoukelli, M.~Mankus, M.~Hasson, M.~Lennie, M.~Reso, M.~Groshev, M.~Naumov, M.~Lathi, M.~Keneally, M.~L. Seltzer, M.~Valko, M.~Restrepo, M.~Patel, M.~Vyatskov, M.~Samvelyan, M.~Clark, M.~Macey, M.~Wang, M.~J. Hermoso, M.~Metanat, M.~Rastegari, M.~Bansal, N.~Santhanam, N.~Parks, N.~White, N.~Bawa, N.~Singhal, N.~Egebo, N.~Usunier, N.~P. Laptev, N.~Dong, N.~Zhang, N.~Cheng, O.~Chernoguz, O.~Hart, O.~Salpekar, O.~Kalinli, P.~Kent, P.~Parekh, P.~Saab, P.~Balaji, P.~Rittner, P.~Bontrager, P.~Roux, P.~Dollar, P.~Zvyagina, P.~Ratanchandani, P.~Yuvraj, Q.~Liang, R.~Alao, R.~Rodriguez, R.~Ayub, R.~Murthy, R.~Nayani, R.~Mitra, R.~Li, R.~Hogan, R.~Battey, R.~Wang, R.~Maheswari, R.~Howes, R.~Rinott, S.~J. Bondu, S.~Datta, S.~Chugh, S.~Hunt, S.~Dhillon, S.~Sidorov, S.~Pan, S.~Verma,
  S.~Yamamoto, S.~Ramaswamy, S.~Lindsay, S.~Lindsay, S.~Feng, S.~Lin, S.~C. Zha, S.~Shankar, S.~Zhang, S.~Zhang, S.~Wang, S.~Agarwal, S.~Sajuyigbe, S.~Chintala, S.~Max, S.~Chen, S.~Kehoe, S.~Satterfield, S.~Govindaprasad, S.~Gupta, S.~Cho, S.~Virk, S.~Subramanian, S.~Choudhury, S.~Goldman, T.~Remez, T.~Glaser, T.~Best, T.~Kohler, T.~Robinson, T.~Li, T.~Zhang, T.~Matthews, T.~Chou, T.~Shaked, V.~Vontimitta, V.~Ajayi, V.~Montanez, V.~Mohan, V.~S. Kumar, V.~Mangla, V.~Albiero, V.~Ionescu, V.~Poenaru, V.~T. Mihailescu, V.~Ivanov, W.~Li, W.~Wang, W.~Jiang, W.~Bouaziz, W.~Constable, X.~Tang, X.~Wang, X.~Wu, X.~Wang, X.~Xia, X.~Wu, X.~Gao, Y.~Chen, Y.~Hu, Y.~Jia, Y.~Qi, Y.~Li, Y.~Zhang, Y.~Zhang, Y.~Adi, Y.~Nam, Yu, Wang, Y.~Hao, Y.~Qian, Y.~He, Z.~Rait, Z.~DeVito, Z.~Rosnbrick, Z.~Wen, Z.~Yang, and Z.~Zhao, ``The llama 3 herd of models,'' 2024. [Online]. Available: \url{https://arxiv.org/abs/2407.21783}
\BIBentrySTDinterwordspacing

\bibitem{emani2023comprehensive}
\BIBentryALTinterwordspacing
M.~Emani, S.~Foreman, V.~Sastry, Z.~Xie, S.~Raskar, W.~Arnold, R.~Thakur, V.~Vishwanath, and M.~E. Papka, ``A comprehensive performance study of large language models on novel ai accelerators,'' 2023. [Online]. Available: \url{https://arxiv.org/abs/2310.04607}
\BIBentrySTDinterwordspacing

\bibitem{fougner2018parameter}
C.~Fougner and S.~Boyd, ``Parameter selection and preconditioning for a graph form solver,'' \emph{Emerging Applications of Control and Systems Theory: A Festschrift in Honor of Mathukumalli Vidyasagar}, pp. 41--61, 2018.

\bibitem{gao2017tetris}
M.~Gao, J.~Pu, X.~Yang, M.~Horowitz, and C.~Kozyrakis, ``Tetris: Scalable and efficient neural network acceleration with 3d memory,'' in \emph{Proceedings of the Twenty-Second International Conference on Architectural Support for Programming Languages and Operating Systems}, 2017, pp. 751--764.

\bibitem{geoffrey2021habitat}
X.~Y. Geoffrey, Y.~Gao, P.~Golikov, and G.~Pekhimenko, ``Habitat: A $\{$Runtime-Based$\}$ computational performance predictor for deep neural network training,'' in \emph{2021 USENIX Annual Technical Conference (USENIX ATC 21)}, 2021, pp. 503--521.

\bibitem{gilbert2024looptree}
M.~Gilbert, Y.~N. Wu, J.~S. Emer, and V.~Sze, ``Looptree: Exploring the fused-layer dataflow accelerator design space,'' \emph{IEEE Transactions on Circuits and Systems for Artificial Intelligence}, 2024.

\bibitem{gomez2015netflix}
\BIBentryALTinterwordspacing
C.~A. Gomez-Uribe and N.~Hunt, ``The netflix recommender system: Algorithms, business value, and innovation,'' \emph{ACM Trans. Manage. Inf. Syst.}, vol.~6, no.~4, dec 2016. [Online]. Available: \url{https://doi.org/10.1145/2843948}
\BIBentrySTDinterwordspacing

\bibitem{hadidi2017demystifying}
R.~Hadidi, B.~Asgari, B.~A. Mudassar, S.~Mukhopadhyay, S.~Yalamanchili, and H.~Kim, ``Demystifying the characteristics of 3d-stacked memories: A case study for hybrid memory cube,'' in \emph{2017 IEEE international symposium on Workload characterization (IISWC)}.\hskip 1em plus 0.5em minus 0.4em\relax IEEE, 2017, pp. 66--75.

\bibitem{huang2021cosa}
\BIBentryALTinterwordspacing
Q.~Huang, M.~Kang, G.~Dinh, T.~Norell, A.~Kalaiah, J.~Demmel, J.~Wawrzynek, and Y.~S. Shao, ``Cosa: Scheduling by constrained optimization for spatial accelerators,'' in \emph{Proceedings of the 48th Annual International Symposium on Computer Architecture}, ser. ISCA '21.\hskip 1em plus 0.5em minus 0.4em\relax IEEE Press, 2021, p. 554–566. [Online]. Available: \url{https://doi.org/10.1109/ISCA52012.2021.00050}
\BIBentrySTDinterwordspacing

\bibitem{mindthegap}
Q.~Huang, P.-A. Tsai, J.~S. Emer, and A.~Parashar, ``Mind the gap: Attainable data movement and operational intensity bounds for tensor algorithms,'' in \emph{2024 ACM/IEEE 51th Annual International Symposium on Computer Architecture (ISCA) (To Appear)}, 2024.

\bibitem{huang2019gpipe}
Y.~Huang, Y.~Cheng, A.~Bapna, O.~Firat, M.~X. Chen, D.~Chen, H.~Lee, J.~Ngiam, Q.~V. Le, Y.~Wu, and Z.~Chen, \emph{GPipe: efficient training of giant neural networks using pipeline parallelism}.\hskip 1em plus 0.5em minus 0.4em\relax Red Hook, NY, USA: Curran Associates Inc., 2019.

\bibitem{isaev2023calculon}
\BIBentryALTinterwordspacing
M.~Isaev, N.~Mcdonald, L.~Dennison, and R.~Vuduc, ``Calculon: a methodology and tool for high-level co-design of systems and large language models,'' in \emph{Proceedings of the International Conference for High Performance Computing, Networking, Storage and Analysis}, ser. SC '23.\hskip 1em plus 0.5em minus 0.4em\relax New York, NY, USA: Association for Computing Machinery, 2023. [Online]. Available: \url{https://doi.org/10.1145/3581784.3607102}
\BIBentrySTDinterwordspacing

\bibitem{jesse2023large}
\BIBentryALTinterwordspacing
K.~Jesse, T.~Ahmed, P.~T. Devanbu, and E.~Morgan, ``Large language models and simple, stupid bugs,'' in \emph{2023 IEEE/ACM 20th International Conference on Mining Software Repositories (MSR)}.\hskip 1em plus 0.5em minus 0.4em\relax Los Alamitos, CA, USA: IEEE Computer Society, may 2023, pp. 563--575. [Online]. Available: \url{https://doi.ieeecomputersociety.org/10.1109/MSR59073.2023.00082}
\BIBentrySTDinterwordspacing

\bibitem{jia2019beyond}
\BIBentryALTinterwordspacing
Z.~Jia, M.~Zaharia, and A.~Aiken, ``Beyond data and model parallelism for deep neural networks.'' in \emph{Proceedings of Machine Learning and Systems}, A.~Talwalkar, V.~Smith, and M.~Zaharia, Eds., vol.~1, 2019, pp. 1--13. [Online]. Available: \url{https://proceedings.mlsys.org/paper_files/paper/2019/file/b422680f3db0986ddd7f8f126baaf0fa-Paper.pdf}
\BIBentrySTDinterwordspacing

\bibitem{jouppi2023tpu}
\BIBentryALTinterwordspacing
N.~Jouppi, G.~Kurian, S.~Li, P.~Ma, R.~Nagarajan, L.~Nai, N.~Patil, S.~Subramanian, A.~Swing, B.~Towles, C.~Young, X.~Zhou, Z.~Zhou, and D.~A. Patterson, ``Tpu v4: An optically reconfigurable supercomputer for machine learning with hardware support for embeddings,'' in \emph{Proceedings of the 50th Annual International Symposium on Computer Architecture}, ser. ISCA '23.\hskip 1em plus 0.5em minus 0.4em\relax New York, NY, USA: Association for Computing Machinery, 2023. [Online]. Available: \url{https://doi.org/10.1145/3579371.3589350}
\BIBentrySTDinterwordspacing

\bibitem{jouppi2021ten}
N.~P. Jouppi, D.~Hyun~Yoon, M.~Ashcraft, M.~Gottscho, T.~B. Jablin, G.~Kurian, J.~Laudon, S.~Li, P.~Ma, X.~Ma, T.~Norrie, N.~Patil, S.~Prasad, C.~Young, Z.~Zhou, and D.~Patterson, ``Ten lessons from three generations shaped google’s tpuv4i : Industrial product,'' in \emph{2021 ACM/IEEE 48th Annual International Symposium on Computer Architecture (ISCA)}, 2021, pp. 1--14.

\bibitem{jung2016parallel}
\BIBentryALTinterwordspacing
J.~Jung, C.~Kobayashi, T.~Imamura, and Y.~Sugita, ``Parallel implementation of 3d fft with volumetric decomposition schemes for efficient molecular dynamics simulations,'' \emph{Computer Physics Communications}, vol. 200, pp. 57--65, 2016. [Online]. Available: \url{https://www.sciencedirect.com/science/article/pii/S0010465515004063}
\BIBentrySTDinterwordspacing

\bibitem{kim2016neurocube}
D.~Kim, J.~Kung, S.~Chai, S.~Yalamanchili, and S.~Mukhopadhyay, ``Neurocube: A programmable digital neuromorphic architecture with high-density 3d memory,'' \emph{ACM SIGARCH Computer Architecture News}, vol.~44, no.~3, pp. 380--392, 2016.

\bibitem{kim2022snuhpl}
\BIBentryALTinterwordspacing
J.~Kim, H.~Kwon, J.~Kang, J.~Park, S.~Lee, and J.~Lee, ``Snuhpl: high performance linpack for heterogeneous gpus,'' in \emph{Proceedings of the 36th ACM International Conference on Supercomputing}, ser. ICS '22.\hskip 1em plus 0.5em minus 0.4em\relax New York, NY, USA: Association for Computing Machinery, 2022. [Online]. Available: \url{https://doi.org/10.1145/3524059.3532370}
\BIBentrySTDinterwordspacing

\bibitem{kim2008technology}
J.~Kim, W.~J. Dally, S.~Scott, and D.~Abts, ``Technology-driven, highly-scalable dragonfly topology,'' in \emph{2008 International Symposium on Computer Architecture}, 2008, pp. 77--88.

\bibitem{kong2023mapzero}
\BIBentryALTinterwordspacing
X.~Kong, Y.~Huang, J.~Zhu, X.~Man, Y.~Liu, C.~Feng, P.~Gou, M.~Tang, S.~Wei, and L.~Liu, ``Mapzero: Mapping for coarse-grained reconfigurable architectures with reinforcement learning and monte-carlo tree search,'' in \emph{Proceedings of the 50th Annual International Symposium on Computer Architecture}, ser. ISCA '23.\hskip 1em plus 0.5em minus 0.4em\relax New York, NY, USA: Association for Computing Machinery, 2023. [Online]. Available: \url{https://doi.org/10.1145/3579371.3589081}
\BIBentrySTDinterwordspacing

\bibitem{lee2021task}
Y.~S. Lee and T.~H. Han, ``Task parallelism-aware deep neural network scheduling on multiple hybrid memory cube-based processing-in-memory,'' \emph{IEEE Access}, vol.~9, pp. 68\,561--68\,572, 2021.

\bibitem{leviathan2023fast}
Y.~Leviathan, M.~Kalman, and Y.~Matias, ``Fast inference from transformers via speculative decoding,'' in \emph{International Conference on Machine Learning}.\hskip 1em plus 0.5em minus 0.4em\relax PMLR, 2023, pp. 19\,274--19\,286.

\bibitem{li2019evaluating}
A.~Li, S.~L. Song, J.~Chen, J.~Li, X.~Liu, N.~R. Tallent, and K.~J. Barker, ``Evaluating modern gpu interconnect: Pcie, nvlink, nv-sli, nvswitch and gpudirect,'' \emph{IEEE Transactions on Parallel and Distributed Systems}, vol.~31, no.~1, pp. 94--110, 2020.

\bibitem{lie2021multi}
S.~Lie, ``Multi-million core, multi-wafer ai cluster,'' in \emph{2021 IEEE Hot Chips 33 Symposium (HCS)}, 2021, pp. 1--41.

\bibitem{lie2022cerebras}
S.~Lie, ``Cerebras architecture deep dive: First look inside the hw/sw co-design for deep learning : Cerebras systems,'' in \emph{2022 IEEE Hot Chips 34 Symposium (HCS)}, 2022, pp. 1--34.

\bibitem{lie2024wafer}
S.~Lie, ``Wafer-scale ai: Enabling unprecedented ai compute performance,'' in \emph{2024 IEEE Hot Chips 36 Symposium (HCS)}.\hskip 1em plus 0.5em minus 0.4em\relax IEEE, 2024.

\bibitem{lin2022building}
Z.~Lin, L.~Feng, E.~K. Ardestani, J.~Lee, J.~Lundell, C.~Kim, A.~Kejariwal, and J.~D. Owens, ``Building a performance model for deep learning recommendation model training on gpus,'' in \emph{2022 IEEE International Symposium on Performance Analysis of Systems and Software (ISPASS)}, 2022, pp. 227--229.

\bibitem{liu2021ai}
\BIBentryALTinterwordspacing
Z.~Liu, R.~A. Roberts, M.~Lal-Nag, X.~Chen, R.~Huang, and W.~Tong, ``Ai-based language models powering drug discovery and development,'' \emph{Drug Discovery Today}, vol.~26, no.~11, pp. 2593--2607, 2021. [Online]. Available: \url{https://www.sciencedirect.com/science/article/pii/S1359644621002816}
\BIBentrySTDinterwordspacing

\bibitem{melchert2023apex}
\BIBentryALTinterwordspacing
J.~Melchert, K.~Feng, C.~Donovick, R.~Daly, R.~Sharma, C.~Barrett, M.~A. Horowitz, P.~Hanrahan, and P.~Raina, ``Apex: A framework for automated processing element design space exploration using frequent subgraph analysis,'' in \emph{Proceedings of the 28th ACM International Conference on Architectural Support for Programming Languages and Operating Systems, Volume 3}, ser. ASPLOS 2023.\hskip 1em plus 0.5em minus 0.4em\relax New York, NY, USA: Association for Computing Machinery, 2023, p. 33–45. [Online]. Available: \url{https://doi.org/10.1145/3582016.3582070}
\BIBentrySTDinterwordspacing

\bibitem{10.1145/3620666.3651335}
\BIBentryALTinterwordspacing
X.~Miao, G.~Oliaro, Z.~Zhang, X.~Cheng, Z.~Wang, Z.~Zhang, R.~Y.~Y. Wong, A.~Zhu, L.~Yang, X.~Shi, C.~Shi, Z.~Chen, D.~Arfeen, R.~Abhyankar, and Z.~Jia, ``Specinfer: Accelerating large language model serving with tree-based speculative inference and verification,'' in \emph{Proceedings of the 29th ACM International Conference on Architectural Support for Programming Languages and Operating Systems, Volume 3}, ser. ASPLOS '24.\hskip 1em plus 0.5em minus 0.4em\relax New York, NY, USA: Association for Computing Machinery, 2024, p. 932–949. [Online]. Available: \url{https://doi.org/10.1145/3620666.3651335}
\BIBentrySTDinterwordspacing

\bibitem{mlperf2024inference}
\BIBentryALTinterwordspacing
{MLCommons}, ``New mlperf inference v4.1 benchmark results highlight rapid hardware and software innovations in generative ai systems,'' August 2024. [Online]. Available: \url{https://mlcommons.org/2024/08/mlperf-inference-v4-1-results/}
\BIBentrySTDinterwordspacing

\bibitem{moolchandani2023amped}
D.~Moolchandani, J.~Kundu, F.~Ruelens, P.~Vrancx, T.~Evenblij, and M.~Perumkunnil, ``Amped: An analytical model for performance in distributed training of transformers,'' in \emph{2023 IEEE International Symposium on Performance Analysis of Systems and Software (ISPASS)}.\hskip 1em plus 0.5em minus 0.4em\relax IEEE, 2023, pp. 306--315.

\bibitem{10.1145/3470496.3533727}
\BIBentryALTinterwordspacing
D.~Mudigere, Y.~Hao, J.~Huang, Z.~Jia, A.~Tulloch, S.~Sridharan, X.~Liu, M.~Ozdal, J.~Nie, J.~Park, L.~Luo, J.~A. Yang, L.~Gao, D.~Ivchenko, A.~Basant, Y.~Hu, J.~Yang, E.~K. Ardestani, X.~Wang, R.~Komuravelli, C.-H. Chu, S.~Yilmaz, H.~Li, J.~Qian, Z.~Feng, Y.~Ma, J.~Yang, E.~Wen, H.~Li, L.~Yang, C.~Sun, W.~Zhao, D.~Melts, K.~Dhulipala, K.~Kishore, T.~Graf, A.~Eisenman, K.~K. Matam, A.~Gangidi, G.~J. Chen, M.~Krishnan, A.~Nayak, K.~Nair, B.~Muthiah, M.~khorashadi, P.~Bhattacharya, P.~Lapukhov, M.~Naumov, A.~Mathews, L.~Qiao, M.~Smelyanskiy, B.~Jia, and V.~Rao, ``Software-hardware co-design for fast and scalable training of deep learning recommendation models,'' in \emph{Proceedings of the 49th Annual International Symposium on Computer Architecture}, ser. ISCA '22.\hskip 1em plus 0.5em minus 0.4em\relax New York, NY, USA: Association for Computing Machinery, 2022, p. 993–1011. [Online]. Available: \url{https://doi.org/10.1145/3470496.3533727}
\BIBentrySTDinterwordspacing

\bibitem{narayanan2021efficient}
\BIBentryALTinterwordspacing
D.~Narayanan, M.~Shoeybi, J.~Casper, P.~LeGresley, M.~Patwary, V.~Korthikanti, D.~Vainbrand, P.~Kashinkunti, J.~Bernauer, B.~Catanzaro, A.~Phanishayee, and M.~Zaharia, ``Efficient large-scale language model training on gpu clusters using megatron-lm,'' in \emph{Proceedings of the International Conference for High Performance Computing, Networking, Storage and Analysis}, ser. SC '21.\hskip 1em plus 0.5em minus 0.4em\relax New York, NY, USA: Association for Computing Machinery, 2021. [Online]. Available: \url{https://doi.org/10.1145/3458817.3476209}
\BIBentrySTDinterwordspacing

\bibitem{nowatzki2018hybrid}
\BIBentryALTinterwordspacing
T.~Nowatzki, N.~Ardalani, K.~Sankaralingam, and J.~Weng, ``Hybrid optimization/heuristic instruction scheduling for programmable accelerator codesign,'' in \emph{Proceedings of the 27th International Conference on Parallel Architectures and Compilation Techniques}, ser. PACT '18.\hskip 1em plus 0.5em minus 0.4em\relax New York, NY, USA: Association for Computing Machinery, 2018. [Online]. Available: \url{https://doi.org/10.1145/3243176.3243212}
\BIBentrySTDinterwordspacing

\bibitem{nowatzki2013general}
\BIBentryALTinterwordspacing
T.~Nowatzki, M.~Sartin-Tarm, L.~De~Carli, K.~Sankaralingam, C.~Estan, and B.~Robatmili, ``A general constraint-centric scheduling framework for spatial architectures,'' in \emph{Proceedings of the 34th ACM SIGPLAN Conference on Programming Language Design and Implementation}, ser. PLDI '13.\hskip 1em plus 0.5em minus 0.4em\relax New York, NY, USA: Association for Computing Machinery, 2013, p. 495–506. [Online]. Available: \url{https://doi.org/10.1145/2491956.2462163}
\BIBentrySTDinterwordspacing

\bibitem{parashar2019timeloop}
A.~Parashar, P.~Raina, Y.~S. Shao, Y.-H. Chen, V.~A. Ying, A.~Mukkara, R.~Venkatesan, B.~Khailany, S.~W. Keckler, and J.~Emer, ``Timeloop: A systematic approach to dnn accelerator evaluation,'' in \emph{2019 IEEE International Symposium on Performance Analysis of Systems and Software (ISPASS)}, 2019, pp. 304--315.

\bibitem{pati2023tale}
S.~Pati, S.~Aga, M.~Islam, N.~Jayasena, and M.~D. Sinclair, ``Tale of two cs: Computation vs. communication scaling for future transformers on future hardware,'' in \emph{2023 IEEE International Symposium on Workload Characterization (IISWC)}.\hskip 1em plus 0.5em minus 0.4em\relax IEEE, 2023, pp. 140--153.

\bibitem{prabhakar2024sambanova}
R.~Prabhakar, ``Sambanova sn40l rdu: Breaking the barrier of trillion+ parameter scale gen ai computing,'' in \emph{2024 IEEE Hot Chips 36 Symposium (HCS)}.\hskip 1em plus 0.5em minus 0.4em\relax IEEE, 2024, pp. 1--24.

\bibitem{prabhakar2021sambanova}
R.~Prabhakar and S.~Jairath, ``Sambanova sn10 rdu:accelerating software 2.0 with dataflow,'' in \emph{2021 IEEE Hot Chips 33 Symposium (HCS)}, 2021, pp. 1--37.

\bibitem{prabhakar2022sambanova}
R.~Prabhakar, S.~Jairath, and J.~L. Shin, ``Sambanova sn10 rdu: A 7nm dataflow architecture to accelerate software 2.0,'' in \emph{2022 IEEE International Solid-State Circuits Conference (ISSCC)}, vol.~65, 2022, pp. 350--352.

\bibitem{radford2019language}
\BIBentryALTinterwordspacing
A.~Radford, J.~Wu, R.~Child, D.~Luan, D.~Amodei, and I.~Sutskever, ``Language models are unsupervised multitask learners,'' 2019. [Online]. Available: \url{https://api.semanticscholar.org/CorpusID:160025533}
\BIBentrySTDinterwordspacing

\bibitem{rashidi2020astra}
S.~Rashidi, S.~Sridharan, S.~Srinivasan, and T.~Krishna, ``Astra-sim: Enabling sw/hw co-design exploration for distributed dl training platforms,'' in \emph{2020 IEEE International Symposium on Performance Analysis of Systems and Software (ISPASS)}, 2020, pp. 81--92.

\bibitem{rezaei2023smart}
S.~H.~S. Rezaei, P.~Z. Moghaddam, and M.~Modarressi, ``Smart memory: Deep learning acceleration in 3d-stacked memories,'' \emph{IEEE Computer Architecture Letters}, 2023.

\bibitem{samajdar2020systematic}
A.~Samajdar, J.~M. Joseph, Y.~Zhu, P.~Whatmough, M.~Mattina, and T.~Krishna, ``A systematic methodology for characterizing scalability of dnn accelerators using scale-sim,'' in \emph{2020 IEEE International Symposium on Performance Analysis of Systems and Software (ISPASS)}.\hskip 1em plus 0.5em minus 0.4em\relax IEEE, 2020, pp. 58--68.

\bibitem{shah2024flashattention3fastaccurateattention}
\BIBentryALTinterwordspacing
J.~Shah, G.~Bikshandi, Y.~Zhang, V.~Thakkar, P.~Ramani, and T.~Dao, ``Flashattention-3: Fast and accurate attention with asynchrony and low-precision,'' 2024. [Online]. Available: \url{https://arxiv.org/abs/2407.08608}
\BIBentrySTDinterwordspacing

\bibitem{shoeybi2019megatron}
\BIBentryALTinterwordspacing
M.~Shoeybi, M.~Patwary, R.~Puri, P.~LeGresley, J.~Casper, and B.~Catanzaro, ``Megatron-lm: Training multi-billion parameter language models using model parallelism,'' 2020. [Online]. Available: \url{https://arxiv.org/abs/1909.08053}
\BIBentrySTDinterwordspacing

\bibitem{simonovic2018cosmological}
\BIBentryALTinterwordspacing
M.~Simonović, T.~Baldauf, M.~Zaldarriaga, J.~J. Carrasco, and J.~A. Kollmeier, ``Cosmological perturbation theory using the fftlog: formalism and connection to qft loop integrals,'' \emph{Journal of Cosmology and Astroparticle Physics}, vol. 2018, no.~04, p. 030, apr 2018. [Online]. Available: \url{https://dx.doi.org/10.1088/1475-7516/2018/04/030}
\BIBentrySTDinterwordspacing

\bibitem{thakur2005optimization}
\BIBentryALTinterwordspacing
R.~Thakur, R.~Rabenseifner, and W.~Gropp, ``Optimization of collective communication operations in mpich,'' \emph{The International Journal of High Performance Computing Applications}, vol.~19, no.~1, pp. 49--66, 2005. [Online]. Available: \url{https://doi.org/10.1177/1094342005051521}
\BIBentrySTDinterwordspacing

\bibitem{touvron2023llama}
\BIBentryALTinterwordspacing
H.~Touvron, T.~Lavril, G.~Izacard, X.~Martinet, M.-A. Lachaux, T.~Lacroix, B.~Rozi{\`e}re, N.~Goyal, E.~Hambro, F.~Azhar, A.~Rodriguez, A.~Joulin, E.~Grave, and G.~Lample, ``Llama: Open and efficient foundation language models,'' \emph{ArXiv}, vol. abs/2302.13971, 2023. [Online]. Available: \url{https://api.semanticscholar.org/CorpusID:257219404}
\BIBentrySTDinterwordspacing

\bibitem{wang2023optimized}
\BIBentryALTinterwordspacing
W.~Wang, M.~Ghobadi, K.~Shakeri, Y.~Zhang, and N.~Hasani, ``How to build low-cost networks for large language models (without sacrificing performance)?'' 2023. [Online]. Available: \url{https://arxiv.org/abs/2307.12169}
\BIBentrySTDinterwordspacing

\bibitem{williams2009roofline}
\BIBentryALTinterwordspacing
S.~Williams, A.~Waterman, and D.~Patterson, ``Roofline: an insightful visual performance model for multicore architectures,'' \emph{Commun. ACM}, vol.~52, no.~4, p. 65–76, apr 2009. [Online]. Available: \url{https://doi.org/10.1145/1498765.1498785}
\BIBentrySTDinterwordspacing

\bibitem{won2024tacos}
W.~Won, M.~Elavazhagan, S.~Srinivasan, S.~Gupta, and T.~Krishna, ``Tacos: Topology-aware collective algorithm synthesizer for distributed machine learning,'' in \emph{2024 57th IEEE/ACM International Symposium on Microarchitecture (MICRO)}.\hskip 1em plus 0.5em minus 0.4em\relax IEEE, 2024, pp. 856--870.

\bibitem{won2021exploring}
\BIBentryALTinterwordspacing
W.~Won, S.~Rashidi, S.~Srinivasan, and T.~Krishna, ``Libra: Enabling workload-aware multi-dimensional network topology optimization for distributed training of large ai models,'' in \emph{2024 IEEE International Symposium on Performance Analysis of Systems and Software (ISPASS)}.\hskip 1em plus 0.5em minus 0.4em\relax IEEE, May 2024. [Online]. Available: \url{http://dx.doi.org/10.1109/ispass61541.2024.00028}
\BIBentrySTDinterwordspacing

\bibitem{yang2021pipemare}
\BIBentryALTinterwordspacing
B.~Yang, J.~Zhang, J.~Li, C.~Re, C.~Aberger, and C.~De~Sa, ``Pipemare: Asynchronous pipeline parallel dnn training,'' in \emph{Proceedings of Machine Learning and Systems}, A.~Smola, A.~Dimakis, and I.~Stoica, Eds., vol.~3, 2021, pp. 269--296. [Online]. Available: \url{https://proceedings.mlsys.org/paper_files/paper/2021/file/9412531719be7ccf755c4ff98d0969dc-Paper.pdf}
\BIBentrySTDinterwordspacing

\bibitem{yuan2024llm}
\BIBentryALTinterwordspacing
Z.~Yuan, Y.~Shang, Y.~Zhou, Z.~Dong, Z.~Zhou, C.~Xue, B.~Wu, Z.~Li, Q.~Gu, Y.~J. Lee, Y.~Yan, B.~Chen, G.~Sun, and K.~Keutzer, ``Llm inference unveiled: Survey and roofline model insights,'' 2024. [Online]. Available: \url{https://arxiv.org/abs/2402.16363}
\BIBentrySTDinterwordspacing

\bibitem{zhang2022full}
\BIBentryALTinterwordspacing
D.~Zhang, S.~Huda, E.~Songhori, K.~Prabhu, Q.~Le, A.~Goldie, and A.~Mirhoseini, ``A full-stack search technique for domain optimized deep learning accelerators,'' in \emph{Proceedings of the 27th ACM International Conference on Architectural Support for Programming Languages and Operating Systems}, ser. ASPLOS '22.\hskip 1em plus 0.5em minus 0.4em\relax New York, NY, USA: Association for Computing Machinery, 2022, p. 27–42. [Online]. Available: \url{https://doi.org/10.1145/3503222.3507767}
\BIBentrySTDinterwordspacing

\bibitem{llmcompass}
H.~Zhang, A.~Ning, R.~B. Prabhakar, and D.~Wentzlaff, ``Llmcompass: Enabling efficient hardware design for large language model inference,'' in \emph{2024 ACM/IEEE 51th Annual International Symposium on Computer Architecture (ISCA) (To Appear)}, 2024.

\bibitem{zhang1989efficient}
X.~Zhang, M.~Mckenna, J.~P. Mesirov, and D.~L. Waltz, ``An efficient implementation of the back-propagation algorithm on the connection machine cm-2,'' in \emph{Proceedings of the 2nd International Conference on Neural Information Processing Systems}, ser. NIPS'89.\hskip 1em plus 0.5em minus 0.4em\relax Cambridge, MA, USA: MIT Press, 1989, p. 801–809.

\bibitem{zhang2021sara}
Y.~Zhang, N.~Zhang, T.~Zhao, M.~Vilim, M.~Shahbaz, and K.~Olukotun, ``Sara: Scaling a reconfigurable dataflow accelerator,'' in \emph{2021 ACM/IEEE 48th Annual International Symposium on Computer Architecture (ISCA)}, 2021, pp. 1041--1054.

\bibitem{zhao2020towards}
Z.~Zhao, W.~Sheng, Q.~Wang, W.~Yin, P.~Ye, J.~Li, and Z.~Mao, ``Towards higher performance and robust compilation for cgra modulo scheduling,'' \emph{IEEE Transactions on Parallel and Distributed Systems}, vol.~31, no.~9, pp. 2201--2219, 2020.

\bibitem{zheng2022alpa}
\BIBentryALTinterwordspacing
L.~Zheng, Z.~Li, H.~Zhang, Y.~Zhuang, Z.~Chen, Y.~Huang, Y.~Wang, Y.~Xu, D.~Zhuo, E.~P. Xing, J.~E. Gonzalez, and I.~Stoica, ``Alpa: Automating inter- and {Intra-Operator} parallelism for distributed deep learning,'' in \emph{16th USENIX Symposium on Operating Systems Design and Implementation (OSDI 22)}.\hskip 1em plus 0.5em minus 0.4em\relax Carlsbad, CA: USENIX Association, Jul. 2022, pp. 559--578. [Online]. Available: \url{https://www.usenix.org/conference/osdi22/presentation/zheng-lianmin}
\BIBentrySTDinterwordspacing

\bibitem{zhong2024distserve}
Y.~Zhong, S.~Liu, J.~Chen, J.~Hu, Y.~Zhu, X.~Liu, X.~Jin, and H.~Zhang, ``$\{$DistServe$\}$: Disaggregating prefill and decoding for goodput-optimized large language model serving,'' in \emph{18th USENIX Symposium on Operating Systems Design and Implementation (OSDI 24)}, 2024, pp. 193--210.

\end{thebibliography}

\end{document}